\documentclass[aps,prd,twocolumn,superscriptaddress,showpacs]{revtex4}
\usepackage{graphicx}
\usepackage{amsmath,amssymb,times}

\newcommand{\bequ}{\begin{equation}}
\newcommand{\eequ}{\end{equation}}
\newcommand{\bea}{\begin{eqnarray}}
\newcommand{\eea}{\end{eqnarray}}

\newcommand{\bfis}[1]{\mbox{\boldmath ${\scriptstyle #1}$}}

\newcommand{\vii}{{\bfis i}}

\newcommand{\vix}{{\bfis x}}

\DeclareSymbolFont{boldletters}{OML}{cmm} {b}{it}
\DeclareSymbolFontAlphabet{\mathbit}{boldletters}
\DeclareMathSymbol{\alpha}{\mathalpha}{letters}{"0B}
\DeclareMathSymbol{\beta}{\mathalpha}{letters}{"0C}
\DeclareMathSymbol{\gamma}{\mathalpha}{letters}{"0D}
\DeclareMathSymbol{\delta}{\mathalpha}{letters}{"0E}
\DeclareMathSymbol{\epsilon}{\mathalpha}{letters}{"0F}
\DeclareMathSymbol{\zeta}{\mathalpha}{letters}{"10}
\DeclareMathSymbol{\eta}{\mathalpha}{letters}{"11}
\DeclareMathSymbol{\theta}{\mathalpha}{letters}{"12}
\DeclareMathSymbol{\iota}{\mathalpha}{letters}{"13}
\DeclareMathSymbol{\kappa}{\mathalpha}{letters}{"14}
\DeclareMathSymbol{\lambda}{\mathalpha}{letters}{"15}
\DeclareMathSymbol{\mu}{\mathalpha}{letters}{"16}
\DeclareMathSymbol{\nu}{\mathalpha}{letters}{"17}
\DeclareMathSymbol{\xi}{\mathalpha}{letters}{"18}
\DeclareMathSymbol{\pi}{\mathalpha}{letters}{"19}
\DeclareMathSymbol{\rho}{\mathalpha}{letters}{"1A}
\DeclareMathSymbol{\sigma}{\mathalpha}{letters}{"1B}
\DeclareMathSymbol{\tau}{\mathalpha}{letters}{"1C}
\DeclareMathSymbol{\upsilon}{\mathalpha}{letters}{"1D}
\DeclareMathSymbol{\phi}{\mathalpha}{letters}{"1E}
\DeclareMathSymbol{\chi}{\mathalpha}{letters}{"1F}
\DeclareMathSymbol{\psi}{\mathalpha}{letters}{"20}
\DeclareMathSymbol{\omega}{\mathalpha}{letters}{"21}
\DeclareMathSymbol{\varepsilon}{\mathalpha}{letters}{"22}
\DeclareMathSymbol{\vartheta}{\mathalpha}{letters}{"23}
\DeclareMathSymbol{\varpi}{\mathalpha}{letters}{"24}
\DeclareMathSymbol{\varrho}{\mathalpha}{letters}{"25}
\DeclareMathSymbol{\varsigma}{\mathalpha}{letters}{"26}
\DeclareMathSymbol{\varphi}{\mathalpha}{letters}{"27}
\DeclareMathSymbol{\Gamma}{\mathalpha}{letters}{"00}
\DeclareMathSymbol{\Delta}{\mathalpha}{letters}{"01}
\DeclareMathSymbol{\Theta}{\mathalpha}{letters}{"02}
\DeclareMathSymbol{\Lambda}{\mathalpha}{letters}{"03}
\DeclareMathSymbol{\Xi}{\mathalpha}{letters}{"04}
\DeclareMathSymbol{\Pi}{\mathalpha}{letters}{"05}
\DeclareMathSymbol{\Sigma}{\mathalpha}{letters}{"06}
\DeclareMathSymbol{\Upsilon}{\mathalpha}{letters}{"07}
\DeclareMathSymbol{\Phi}{\mathalpha}{letters}{"08}
\DeclareMathSymbol{\Psi}{\mathalpha}{letters}{"09}
\DeclareMathSymbol{\Omega}{\mathalpha}{letters}{"0A}

\begin{document}
%\preprint{SAGA-HE-???}
\title{Nonanalyticity, sign problem and Polyakov line in $Z_3$-symmetric heavy quark model at low temperature: Phenomenological model analyses} 

\author{Hiroaki Kouno}
\email[]{kounoh@cc.saga-u.ac.jp}
\affiliation{Department of Physics, Saga University,
             Saga 840-8502, Japan}

\author{Kouji Kashiwa}
\email[]{Kashiwa@fit.ac.jp}
\affiliation{Fukuoka Institute of Technology, Wajiro, Fukuoka 811-0295, Japan}

\author{Takehiro Hirakida}
\email[]{hirakida@izumi.ac.jp}
\affiliation{Izumi Chuo high school, 
             Izumi 899-0213, Japan}

\date{\today}

\begin{abstract}
The nonanalyticity and the sign problem in the $Z_3$-symmetric heavy quark model at low temperature are studied phenomenologically. 
For the free heavy quarks, the nonanalyticity is analyzed in the relation to the zeros of the grand canonical partition function.  
The $Z_3$-symmetric effective Polyakov-line model (EPLM)  in strong coupling limit is also considered as an phenomenological model of $Z_3$-symmetric QCD with large quark mass at low temperature. 
We examine how the $Z_3$-symmetric EPLM approaches to the original one in the zero-temperature limit. 
The effects of the $Z_3$-symmetry affect the structure of zeros of the microscopic probability density function at the nonanalytic point.  
The average value of the Polyakov line can detect the structure, while the other thermodynamic quantities are not sensible to the structure in the zero-temperature limit.  
The effect of the imaginary quark chemical potential is also discussed. 
The imaginary part of the quark number density is very sensitive to the symmetry structure at the nonanalytical point.
For a particular value of the imaginary quark number chemical potential, large quark number may be induced in the vicinity of the nonanalytical point. 
\end{abstract}

\maketitle

%%%%%%%%%%%%%%%%%%%%%%%%%%%%%%%%%%%%%%%%%%%%%%%%%%%%%%%%%%%%%%%%%%%%%%%%%%%
%%%%%  Introduction 
%%%%%%%%%%%%%%%%%%%%%%%%%%%%%%%%%%%%%%%%%%%%%%%%%%%%%%%%%%%%%%%%%%%%%%%%%%%
\section{Introduction}
\label{sec:intro}
%%%%%%%%%%%%%%%%%%%%%%%%%%%%%%%%%%%%%%%

Study of the quantum chromodynamics (QCD) phase structure at finite temperature $T$ and quark chemical potential $\mu$ is one of the most important 
subjects in particle and nuclear physics, astrophysics and cosmology. 
Nowadays, the first-principle nonperturbative calculation, the lattice QCD (LQCD) simulation has been almost established at $\mu =0$.   
However, for $\mu \neq 0$, LQCD has a famous sign problem and is very difficult to be done correctly.  
An effective action obtained after the integration over the quark fields is complex and the numerical simulation such as the Monte Carlo simulation is very difficult, since we can not construct a proper probability density function. 
Several methods were proposed so far to circumvent the sign problem; namely, 
the reweighting method~\cite{Fodor},
the Taylor expansion method~\cite{Allton,Ejiri_density}
, the analytic continuation from imaginary $\mu$
to real $\mu$~\cite{FP,D'Elia,D'Elia3,FP2010,Nagata,Takahashi}, the complex Langevin simulation
\cite{Aarts_CLE_1,Aarts_CLE_2,Aarts_CLE_3,Sexty,Aarts_James,Greensite:2014cxa}
, the Picard-Lefschetz thimble theory \cite{Aurora_thimbles,Fujii_thimbles,Tanizaki,Tanizaki_2}, and 
the path optimization method~\cite{Mori:2017nwj,Mori:2017pne}.  
Particularly for the case of $ \mu/T >1$ , our understanding of the QCD phase diagram is still far from perfection. 

It was also suggested that the sign problem may be weaker in the $Z_3$-symmetrized QCD than the original one~\cite{Kouno:2015sja}.  
Due to the effects of dynamical quark, the $Z_3$ symmetry which exists in the pure gluon theory and is related to the quark confinement is explicitly broken. 
However, in the symmetric three flavor QCD, the $Z_3$ symmetry can be restored by introducing imaginary isospin chemical potential with the absolute value $i{2\over{3}}\pi T$. 
In this paper, we call the $Z_3$-symmetric QCD "$Z_3$-QCD"~\cite{Hasenfratz:1991ax, Kouno:2012dn_2,Sakai:2012ika,Kouno:2013zr,Kouno:2013mma,Kouno:2015sja,Iritani:2015ara,Cherman:2016hcd,Liu:2016yij,Tanizaki:2017mtm,Cherman:2017tey}. 
In $Z_3$-QCD and its effective models, the sign problem is expected to be weaker than the original ones, since the number of configurations of which the effective action are real  increases by symmetrizing the theories.   
In fact, the $Z_3$-symmetric three-dimensional three state Potts model has no sign problem~\cite{Hirakida:2016rqd}.  
In the $Z_3$-symmetric effective Polyakov-line model (EPLM), the sign problem remains, but it is much weaker than in EPLM without $Z_3$-symmetry~\cite{Hirakida:2017bye}.  
(In this paper, we call the $Z_3$-symmetric EPLM "$Z_3$-EPLM". )

\begin{figure}[h]
\centering
{\includegraphics[width=0.35\textwidth]{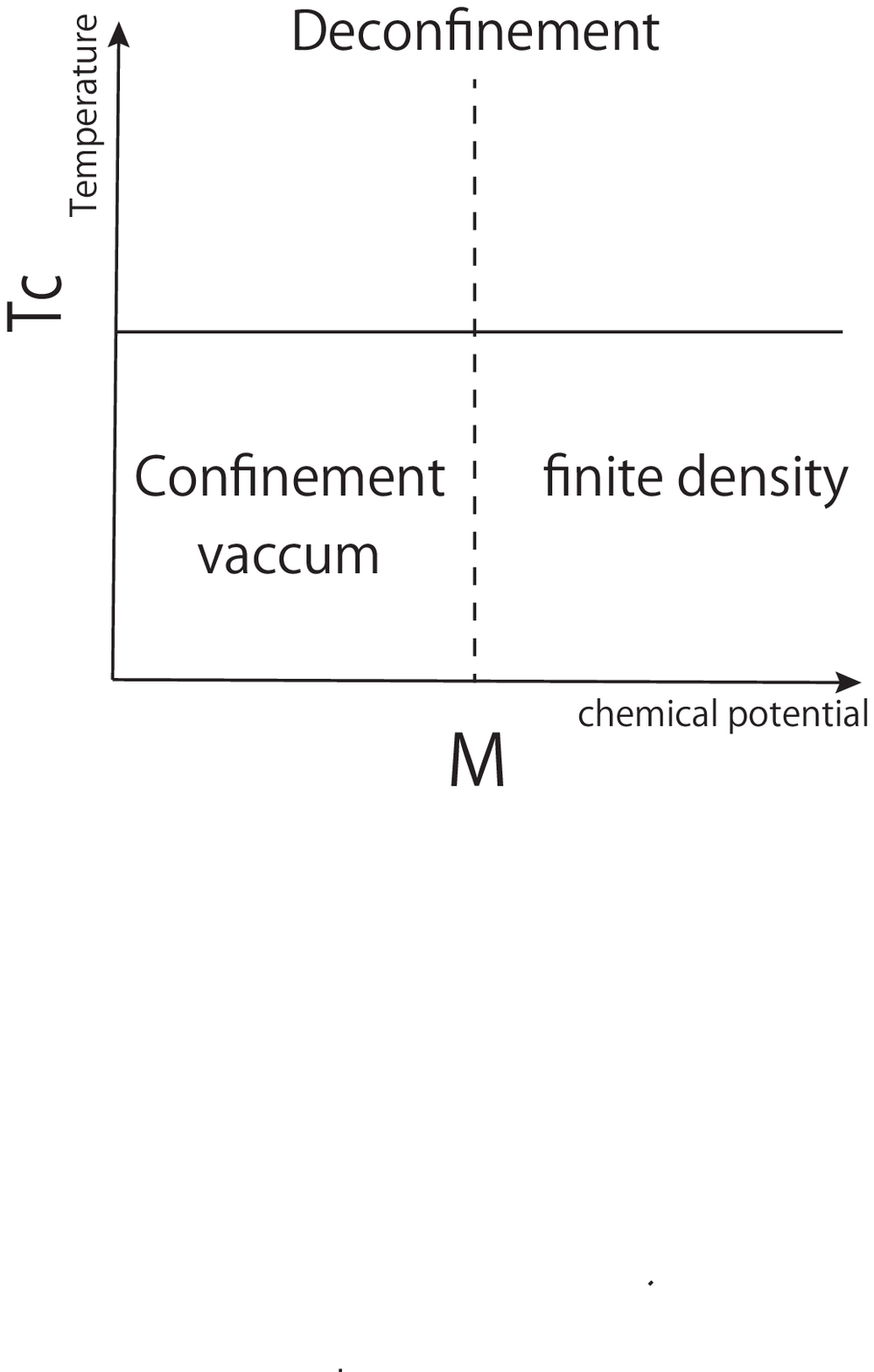}}
\caption{Schematic phase0diagram of $Z_3$-EPLM with quark mass $M$. 
 }
 \label{phase-diagram}
\end{figure}

Figure~\ref{phase-diagram} shows the schematic phase diagram obtained by using $Z_3$-EPLM with the reweighting method~\cite{Hirakida:2017bye}. 
This diagram is also consistent with the one obtained by using the $Z_3$-symmetric three-dimensional three-state Potts model with no sign problem~\cite{Hirakida:2016rqd}. 
In $Z_3$-EPLM, the sign problem happens only nearby the line $\mu =M$ at low temperature. 
Since EPLM is the model of heavy quarks, the chiral symmetry restoration is not expected.  
Hence, it is naturally expected that this sign problem is simply related to the Fermi sphere formation at $\mu =M$. 
In this meaning, here we call this sign problem "a trivial sign problem". 
To detect anomalous phenomena, we need to remove or weaken it.   

%%%%%%%%%%%%%%%%
%%%%%%%%%%%%%%%

It is well known that the imaginary chemical potential is transformed into the change of the temporal boundary condition of quark fields by redefining  the quark fields~\cite{Roberge:1986mm}. 
The temporal boundary condition is irrelevant in the zero temperature limit, namely $\beta =1/T \to \infty$. 
Hence, $Z_3$-QCD (and its effective model) approaches to the original QCD (and effective model). 
In fact, in Ref.~\cite{Kouno:2015sja}, it was shown that the phase diagram of the $Z_3$-symmetric Polyakov-loop extended Nambu-Jona-Lasinio (PNJL) model coincides with the one in the original PNJL model~\cite{Meisinger:1995ih,Dumitru:2002cf,Fukushima:2003fw,Ratti:2005jh,Megias:2004hj} in the zero-temperature limit.  
However, this limit may be nontrivial at finite $\mu$ since nonanalyticity occurs at zero temperature due to the Fermion sphere formation and this phenomena itself is also related to the change of the boundary condition and the zeros of the grand canonical partition function. 
Furthermore, in $Z_3$-symmetric theory, the expectation value of the Polyakov line (loop) vanishes due to the exact $Z_3$-symmetry, while it can be finite in the original model.  
It is nontrivial whether the Polyakov line coincides or not in two models in the zero-temperature limit. 
(Note that the expectation values of the absolute value of the spatial average of the Polyakov line can be finite in $Z_3$-symmetric model and is used to analyze the confinement-deconfinement transition~\cite{Hirakida:2016rqd,Hirakida:2017bye}. )

In this paper, using the heavy quark model, we study  phenomenologically how the $Z_3$-symmetric model approaches to the original one and weaken the sign problem.  
We also examine how the $Z_3$-symmetry affects the nonanalyticity at zero temperature. 
This paper is organized as follows. 
In Sec.~\ref{FHF}, we study the analyticity in $Z_3$-symmetric heavy quark model in the relation to the zeros of the grand canonical partition function $Z$~ \cite{Yang:1952be,Lee:1952ig} and the boundary condition of the quark field. 
The zeros structure of $Z$ of the free lighter fermion gas with spatial momentum is also discussed. 
In Sec.~\ref{EPLM}, using $Z_3$-EPLM in the strong coupling limit, we examine how $Z_3$-EPLM approaches to the original EPLM and weaken the sign problem in the zero temperature limit. 
Relation among the $Z_3$ symmetry, the zeros of the probability density function and the sign problem is discussed. 
It is shown that the Polyakov line at the nonanalytic point can detect the symmetry structure of the zeros, while the other quantities are not sensible to the structure.  
The effect of the imaginary quark chemical potential and the Roberge-Weiss (RW) periodicity~\cite{Roberge:1986mm} at low temperature limit is also discussed. 
Section~\ref{summary} is devoted to a summary.

%%%%%%%%%%%%%%%%%%%%%%%%%%%%%%%%%%%%%%%%%%%%%%%%%%%%%%%%
%%%% free heavy fermion system 
%%%%%%%%%%%%%%%%%%%%%%%%%%%%%%%%%%%%%%%%%%%%%%%%%%%%%%%%
\section{Nonanalyticity in free heavy quark model at zero temperature}
\label{FHF}
%%%%%%%%%%%%%%%%%%%%%%%%%%%%%%%%%%%%%%%%%%%%%%%%%%%%%%%%

\subsection{EOS of Free fermion at zero-temperature}
\label{FF0}

In this subsection, we briefly summarize the nonanalyticity of the equation of states (EOS) of the free fermion at $T=0$. 
For the free fermion with finite mass $M$, thermodynamical quantities vanish at $T=0$, when $\mu <M$. 
When $\mu \ge M$,    
the pressure $P$ of the free fermion gas at $T=0$ is given by 
%%%%%%%%%%%%%%
\begin{eqnarray}
P&=&{g\mu\over{6\pi^2}}(\mu^2-M^2)^{3/2}
\nonumber\\
&&
-{g\over{8\pi^2}}\left\{\mu \sqrt{\mu^2-M^2}\left(\mu^2-{M^2\over{2}}\right)\right. 
\nonumber\\
&&\left. -{1\over{2}}M^4\log{\left({\mu+\sqrt{\mu^2-M^2}\over{M}}\right)} \right\}, 
\label{pressure0}
\end{eqnarray}
%%%%%%%%%%%%%%
where $g$ is the fermion degree of freedom including the number of spin states. 
The fermion number density and its derivative with respect to $\mu$ are given by  
%%%%%%%%%%%%%%%%%%
\begin{eqnarray}
\rho={\partial P\over{\partial \mu}}={g\over{6\pi^2}}
(\mu^2 -m^2)^{3/2}, 
\label{numberdensity0}
\end{eqnarray}
%%%%%%%%%%%%%%%%%%
and 
%%%%%%%%%%%%%%%%%%
\begin{eqnarray}
{\partial\rho\over{\partial \mu}}={\partial^2 P\over{\partial \mu^2}}={g\mu\over{4\pi^2}}\sqrt{\mu^2-m^2}. 
\label{dnumber0}
\end{eqnarray}
%%%%%%%%%%%%%%%%%
Note that these quantities are continuous at $\mu =M$. 
However, the third derivative of the pressure  
%%%%%%%%%%%%%%%%%
\begin{eqnarray}
{\partial^2 \rho \over{\partial \mu^2}}={\partial^3 P\over{\partial \mu^3}}={g\over{4\pi^2}}\left( \sqrt{\mu^2-m^2}+{\mu^2\over{\sqrt{\mu^2-m^2}}}\right), 
\label{d2number0}
\end{eqnarray}
%%%%%%%%%%%%%%%%%
is divergent at $\mu \to M+0$. 
Hence the pressure is nonanalytic at $\mu =M$.

\subsection{Free heavy quark model}
\label{FHFL}

In this subsection, we consider the free heavy quark model (FHQM) on the lattice with $N_{\rm s}$, $N_f=3$ and $N_c=3$ where $N_{\rm s}$, $N_f$ and $N_c$ are the number of the spatial sites, the number of flavor and the number of  color , respectively.   
Quarks in the heavy mass limit have no spatial momentum and the energy of them is always equal to their mass $M_f~(f=u,d,s)$.   
Hence, the grand canonical partition function is given by 
%%%%%%%%%%%%%%%%%%%%%
\begin{eqnarray}
Z(T,\mu)
&=&\prod_{f=u,d,s}(1+e^{\beta (\mu_f -M_f)})^N
\nonumber\\
\times 
&&(1+e^{\beta (-\mu_f-M_f)})^N,  
\label{grandcanonical}
\end{eqnarray}
%%%%%%%%%%%%%%%%%%%%%
where $N=2N_cN_s$, and $M_f$ and $\mu_f$ are the mass and the chemical potential for the $f$ quark.   
In (\ref{grandcanonical}), the antiquark contributions are included since they are important for the reality of $Z$ when imaginary chemical potential is introduced, although the contributions vanish in the limit $M_f, \mu_f \to \infty$.    
 If we put $\mu_f =M_f+i(2k+1)\pi T$ with an integer $k$, we obtain 
$Z=0$. 
When $T$ approaches to zero, 
the location of the zeros of $Z$ approaches a real value $\mu_f =M_f$. 
Hence, in the analogy of the famous Lee-Yang theorem~\cite{Yang:1952be, Lee:1952ig}, the (dimensionless) pressure
%%%%%%%%%%%%%%%%%%%%%%
\begin{eqnarray}
P=\lim_{N_{\rm s}\to \infty}{1\over{\beta N_s}}\log{Z}
\label{pressure}
\end{eqnarray}
%%%%%%%%%%%%%%%%%%%%%
is expected to be non-analytic at $\mu=M_f$ when $\beta \to \infty$. 
(For the application of the Lee-Yang theorem to the QCD phase transitions, e.g., see~\cite{Nagata:2014fra} and references therein. )

Note that, beside the infinite limit of the spatial volume $N_{\rm s}$, here we also take the infinite limit of the imaginary time ($\tau$) length $\beta$. 
It is known that,  by the redefinition of the quark filed
%%%%%%%%%%%%%%%%%%%%%
\begin{eqnarray}
q_f\to e^{i\theta T\tau}q_f, 
\label{redefqf}
\end{eqnarray}
%%%%%%%%%%%%%%%%%%%%%
the imaginary chemical potential $\mu_{\rm I}=i\theta T$ can be transformed into the twisted temporal boundary condition 
%%%%%%%%%%%%%
\begin{eqnarray}
q_f(\tau =\beta )=-e^{-i\theta}q_f(\tau =0). 
\label{twisted} 
\end{eqnarray}
%%%%%%%%%%%%%%
Hence, if $\theta =(2k+1)\pi$, the quark boundary condition is a periodic boundary condition. 
It should be noted that, except for the singular point $\mu =M_f+i(2k+1)\pi$, the twisted model approaches to the original model in the limit $\beta \to \infty$, since the boundary condition becomes irrelevant in this limit. 
(Since $\theta$ has a trivial periodicity $2\pi$, in this paper, we restrict $\theta$ in the region $(-\pi, \pi ]$ for simplicity. )

Hereafter, we consider the flavor symmetric case, namely, $M_u=M_d=M_s=M$, unless otherwise mentioned.  
The derivative of $P$ with respect to $\mu$, namely, the (dimensionless) number density is given by 
%%%%%%%%%%%%%%%
\begin{eqnarray}
n_{\rm q}  &=&\lim_{N_{\rm s}\to \infty} {1\over{\beta N_{\rm s}}}{\partial \log{Z}\over{\partial \mu}}
\nonumber\\
&=&2N_fN_c\left\{ {1\over{e^{\beta (M-\mu )}+1}}
+{1\over{e^{\beta (M+\mu)}+1}}\right\}
\nonumber\\
&=&
\left\{ 
\begin{array}{cc}
0 &(\mu <M,~~~\beta \to \infty) \\
N_fN_c & (\mu = M,~~~\beta \to \infty ) \\
2N_fN_c & (\mu > M,~~~\beta \to \infty ) 
\end{array}
\right. , 
\label{numberdensity}
\end{eqnarray}
%%%%%%%%%%%%%%%%
and it is clear that $n_{\rm q}$ is nonanalytic at $\mu=M$ and $T=0$. 

\begin{figure}[h]
%\centering
%\begin{center}
\centerline{ 
\includegraphics[width=0.40\textwidth]{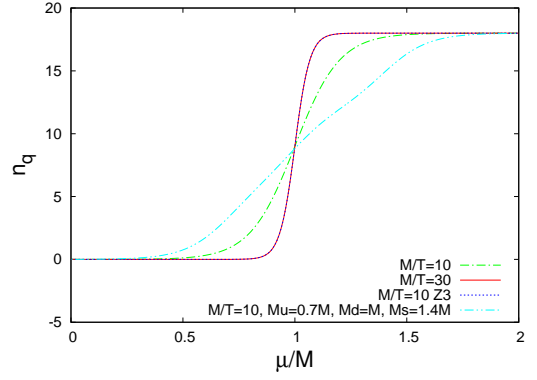}}
%\vspace{7mm}
\caption{The $\mu$-dependence of the quark number density $n_{\rm q}$. 
The dash-dotted and solid lines represent the results in FHQM with $M/T=10,30$, respectively. 
The dotted line represents the result in $Z_3$-FHQM with $M/T=10$ and coincides with the solid line.  
The dash-dot-dotted line represents the result in FHQM with $M_u=0.7M$, $M_d=M$ and $M_s=1.4M$ ($M/T=10$). }
\label{free_n}
%\end{center}
\end{figure}

\begin{figure}[h]
%\centering
\centerline{
\includegraphics[width=0.40\textwidth]{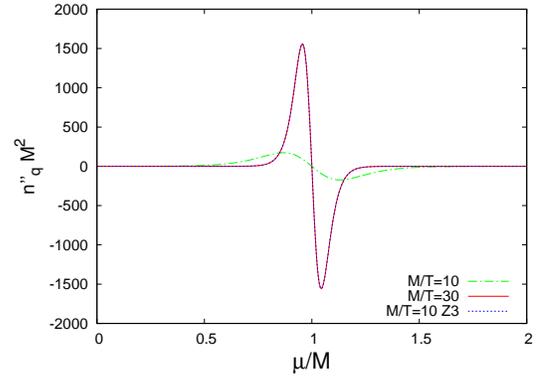}}
%\vspace{7mm}
 \caption{The $\mu$-dependence the second derivative $n_{\rm q}^{\prime\prime}$ of the quark number density with respect to the quark chemical potential.  
Note that $n_{\rm q}^{\prime\prime}$ is multiplied by the factor $M^2$ and dimensionless. 
The dash-dotted and solid lines represent the results in FHQM with $M/T=10,30$, respectively. 
The dotted line represents the result in $Z_3$-FHQM with $M/T=10$ and coincides with the solid line.   
 }
 \label{free_d2n}
\end{figure}

However, in the actual numerical simulations such as EPLM, we can not put ${\beta }=\infty$.  
Hence, $n_{\rm q}$ and its derivatives are continuous functions of $\mu$ as seen in Figs.~\ref{free_n} and \ref{free_d2n}. 
Near the point $\mu =M$, the quark number density $n_{\rm q}$ increases monotonically. The second derivative $n_{\rm q}^{\prime\prime}={\partial^2 n_{\rm q}\over{\partial \mu^2}}$ has maximum and minimum around the point $\mu =M$. 
When $M/T$ becomes larger, $n_{\rm q}$ increases more rapidly and the absolute values of the maximum and minimum of $n^{\prime\prime}$ becomes larger, while the width of the peaks becomes narrower. 

%%%%%%%%%%%%%%
%%%%%%%%%%%%%%%

%%%%%%%%%%%%%%
%%%%%%%%%%%%%%%

\begin{figure}[h]
%\centering
 %\begin{center}
\centerline{
\includegraphics[width=0.40\textwidth]{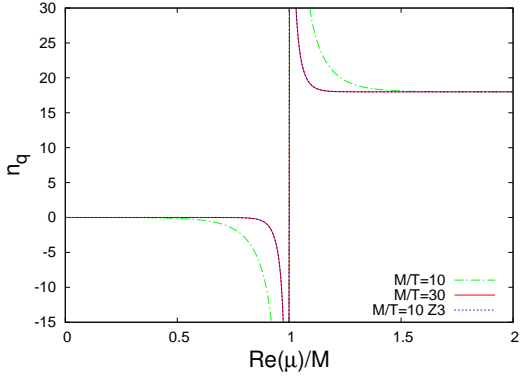}}
%\vspace{7mm}
 \caption{The ${\rm Re}(\mu )$-dependence of the quark number density $n_{\rm q}$ when $\theta =\pi$.  
 The dash-dotted and solid lines represent the results in FHQM with $M/T=10,30$, respectively. 
The dotted line represents the result in $Z_3$-FHQM with $M/T=10$ and coincides with the solid line. 
 }
 \label{free_n_pi}
%\end{center}
\end{figure}

\begin{figure}[h]
%\centering
\centerline{
\includegraphics[width=0.40\textwidth]{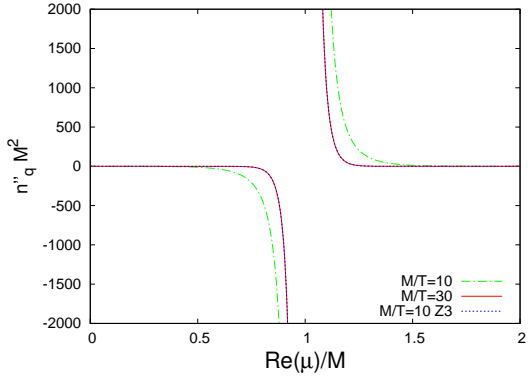}}
%\vspace{7mm}
 \caption{The ${\rm Re}(\mu )$-dependence of the second derivative $n_{\rm q}^{\prime\prime}$ of the quark number density when $\theta =\pi$. The dash-dotted and solid lines represent the results with $M/T=10,30$, respectively. 
The dotted line represent the result in $Z_3$-FHQm with $M/T=10$ and coincides the solid line. 
}
 \label{free_d2n_pi}
\end{figure}

Figs.~\ref{free_n_pi} and \ref{free_d2n_pi} are the same as Figs.~\ref{free_n} and \ref{free_d2n}, respectively, but for $\theta =\pi$.  
As ${\rm Re}(\mu )$ approaches to $M$, $n_{\rm q}$ and $n_{\rm q}^{\prime\prime}$ diverge. 
The region of the divergent behavior be narrower as $M/T$ be larger. 
Hence, it is expected that the results with $\theta =\pi$ approach to those with $\theta =0$ except for the point of ${\rm Re}(\mu )=M$. 
Here, we only show the results of the odd derivatives of the pressure $P$ with respect to the chemical potential $\mu$. 
As is seen in the next section, in EPLM, these odd derivatives are related to the sign problem around $\mu =M$.

%%%%%%%%%%%%%%
%%%%%%%%%%%%%%%

%%%%%%%%%%%%%%
%%%%%%%%%%%%%

As is seen in the previous subsection, in the case of the free fermion with smaller mass and spatial momentum, the number density itself is continuous at $\mu =M$. 
It seems that the effects of the spatial momentum make the transition smoother.  
At finite temperature, the pressure of the free fermion gas is given by
%%%%%%%%%%%%%%%%%%%%%%
\begin{eqnarray}
P&=&{g\over{4\pi^2}}\left(  \int_0^\infty~dp~p^2~T\log{(1+e^{-\beta (E-\mu)} )}\right. 
\nonumber\\
&&\left. + \int_0^\infty~dp~p^2~T\log{(1+e^{-\beta (E+\mu)} )}\right). 
\label{FHF_P0}
\end{eqnarray}
%%%%%%%%%%%%%%%%%%%%%
Then, the grand canonical partition function is given by 
%%%%%%%%%%%%%%%%%%%%%
\begin{eqnarray}
Z&=&\lim_{\Lambda\to\infty} \lim_{N\to \infty}\prod_{n=0}^N (1+e^{-\beta (\sqrt{(n\Delta p)^2+M^2}-\mu)})^{gn^2\Delta p^3\over{4\pi^2}}
\nonumber\\
&&\times  (1+e^{-\beta (\sqrt{(n\Delta p)^2+M^2}+\mu)})^{gn^2\Delta p^3 \over{4\pi^2}}, 
\label{FHF_Z}
\end{eqnarray}
%%%%%%%%%%%%%%%%%%%
where $\Delta p={\Lambda\over{N}}$. 
The function $(1+e^{-\beta (\sqrt{(n\Delta p)^2+M^2}-\mu)})$ is zero at $\mu =\sqrt{(n\Delta p)^2+M^2}+i\pi$.
Zeros of  $(1+e^{-\beta (\sqrt{(n\Delta p)^2+M^2}-\mu)})$ depend on the absolute value $p=n\Delta p$ of the spatial momentum of fermions. 
The set of zeros of $Z$ forms a continuous structure. 
This structure of zeros of $Z$ may smoothen the transition. 
In Fig.~\ref{free_n}, the quark number density $n_{\rm q}$ in FHQM with $M_u=0.5M$, $M_d=M$ and $M_s=1.5M$ is shown for $M/T=10$.  
We see that $n_{\rm q}$ increases slowly as $\mu$ increases. 
For the infinite flavor quarks with mass $\sqrt{(n\Delta p)^2+M^2}$ with infinitesimal degree of freedom, ${gn^2\Delta p^3\over{4\pi^2}}$, it can be expected that $n_{\rm q}$ increases slowly, even if the limit $M/T\to \infty$ is taken. 
In this meaning, the result in FHQM with nonsymmetric flavors mimics the free quark model with lighter mass and the spatial momentum.

%%%%%%%%%%%%%%%%%%%%%%%%%%%%%%%%%%%%%%%%%%%%%%%%%%%%%%%%%%%%%%%%%%%%%%
\subsection{$Z_3$-symmetrization}
\label{Z3symmetry}
%%%%%%%%%%%%%%%%%%%%%%%%%%%%%%%%%%%%%%%%%%%%%%%%%%%%%%%%%%%%%%%%%%%%%%

In the symmetric three flavor quark model, we consider the case where 
%%%%%%%%%%%%%%%%%%%%%%%%%%%%%%%%%%%%
\begin{eqnarray}
\mu_f=\mu +i\theta_fT ~~~~~(f=u,d,s) 
\label{Z3symmetrization}
\end{eqnarray}
%%%%%%%%%%%%%%%%%%%%%%%%%%%%%%%%%%%% 
with $\theta_u=i{2\pi \over{3}}$, $\theta_d=-i{2\pi\over{3}}$, $\theta_s=0$.  
In this paper, we call this setting "$Z_3$-symmetrization" and $Z_3$-symmetric FHQM "$Z_3$-FHQM".  
Since the additional imaginary chemical potential $i\theta_fT$ is the isospin chemical potential rather than the quark chemical potential, it can not be included in the definition of $\mu$. 
It should be noted that $Z$ is real at $\mu_f=\mu +i\theta_fT$ for real $\mu$ and is not zero at $\mu_f=M+i\theta_f$. 

The setting (\ref{Z3symmetrization}) of chemical potential is related to the so-called $Z_3$ symmetry. 
In fact, by the redefinition of the quark field $q_f$, the imaginary part of the chemical potential $\mu_f$ can be transformed into the temporal ($\tau$) boundary condition 
%%%%%%%%%%%%%%%%%%%%%%%%%%%%%%%%%%%%
\begin{eqnarray}
q_f(\tau =\beta )=-e^{-i\theta_f}~q_f(\tau=0 )~~~~~(f=u,d,s), 
\label{Z3boundary}
\end{eqnarray}
%%%%%%%%%%%%%%%%%%%%%%%%%%%%%%%%%%%%
where $e^{-i\theta_f}$ is an element of the $Z_3$ group. 
In QCD, the $Z_3$-transformation changes the boundary condition of quark field by the factor of the $Z_3$ group element. 
Hence, the $Z_3$ symmetry which exists in the pure gluon theory and related to the quark confinement is explicitly broken. 
However,  in the symmetric three flavor QCD, the $Z_3$ symmetry is restored by use of  the $Z_3$-symmetrization (\ref{Z3symmetrization})~\cite{Kouno:2012dn_2}. 
It should be noted that the $Z_3$-symmetric theory is expected to approach to the original one in the limit $\beta \to \infty$, since the boundary condition is not relevant in the limit. 

In $Z_3$-FHQM, the partition function $Z$ becomes zero at following three points,  
%%%%%%%%%%%%%%%%%%%%%%%%%%%%%%%%%%%%
\begin{eqnarray}
\mu=M -i{\pi \over{3}} T,~~~~~M+i{\pi\over{3}} T,~~~~~M+i\pi T. 
\label{Z3_zero}
\end{eqnarray}
%%%%%%%%%%%%%%%%%%%%%%%%%%%%%%%%%%%% 
Hence there are three zeros of $Z$ in the complex $\mu$ plane.  
However, these zeros correspond to the same nonanalyticity at $\mu =M$ when  the zero temperature limit is taken. 
Therefore, $Z_3$-symmetric FHQM has the same information of the nonanalyticity of the original FHQM at zero temperature, although the two models are different each other at finite temperature. 

It is known that the $Z_3$-symmetrization enhances the confinement-like structure and a quark behaves as the particle with the mass $3M$ rather than that with $M$. 
Hence, the effects of the $Z_3$-symmetrization resembles the ones of the increases of $M/T$. 
In fact, the partition function of $Z_3$-FHQM is given by 
%%%%%%%%%%%%%%%%%%%%%
\begin{eqnarray}
Z_{Z_3}(T,\mu)
&=&(1+e^{(3\mu-3M)/T})^N
 (1+e^{(-3\mu-3M)/T})^N
\nonumber\\
&=&Z(T,3\mu; 3M, N_f=1). 
\label{Z3grandcanonical}
\end{eqnarray}
%%%%%%%%%%%%%%%%%%%%%
Hence, $Z_3$-FHQM has the same properties as the ordinary FHQM with one flavor, threefold mass and threefold chemical potential.  
If we define the baryonic chemical potential $\mu_{\rm B}=3\mu$, the zero of $Z$ locates only at $\mu_{\rm B}=3M+i\pi$. 
The quark number density is given by
%%%%%%%%%%%%%%%%%%%%%
\begin{eqnarray}
n_{{\rm q},Z_3}(T,\mu )
&=&{3\over{N_{\rm s}}}{\partial \log{ Z(T,\mu_{\rm B}; 3M, N_f=1)}\over{\partial \mu_{\rm B}}}
\nonumber\\
&&=n_{\rm q}(T,3\mu ; 3M,N_f=3). 
\label{nq_relation}
\end{eqnarray}
%%%%%%%%%%%%%%%%%%%%
The factor 3 complements the flavor decreasing from 3 to 1 in Eq.~(\ref{Z3grandcanonical}).  
Hence, the number density in $Z_3$-FHQM coincides the one in the ordinary FHQM with three flavor, threefold mass and threefold chemical potential. 
The similar correspondence is seen in $n^{\prime\prime}M^2={\partial^2 n_{\rm q}\over{\partial (\mu/M)^2}}$.    
In Figs.~\ref{free_n}$\sim$\ref{free_d2n_pi}, the ${\rm Re}(\mu )$-dependence of $n_{\rm q}$ and $n_{\rm q}^{\prime\prime}$ of the $Z_3$-FHQM are shown for $\theta =0$ and $\pi$.  
In all case, the perfect coincidence between $Z_3$-FHQM with $M/T=10$ and FHQM with $M/T=30$ is seen.    
When $M/T$ is fixed and $\theta =0$, $n_{\rm q}$ increases more rapidly and the absolute values of the maximum and minimum of $n^{\prime\prime}$ becomes larger in $Z_3$ FHQM than in FHQM, while the width of finite $n^{\prime\prime}$ becomes narrower. 
The localization of the peaks of the odd derivatives makes the sign problem weaker than that in EPLM.

%%%%%%%%%%%%%%%%%%%%%%%%%%%%%%%%%%%%%%%%%%%%%%%%%%%%%%%%
%%%% effective Polyakov-line model
%%%%%%%%%%%%%%%%%%%%%%%%%%%%%%%%%%%%%%%%%%%%%%%%%%%%%%%%
\section{Effective Polyakov-line model at zero temperature}
\label{EPLM}
%%%%%%%%%%%%%%%%%%%%%%%%%%%%%%%%%%%%%%%%%%%%%%%%%%%%%%%%
\subsection{Effective Polyakov-line model }
\label{EPLM_sub}

The grand canonical partition function of EPLM in temporal gauge is given by~\cite{Aarts_James,Greensite:2014cxa,Hirakida:2017bye}
%%%%%%%%%%%%%%
\begin{eqnarray}
Z&=&\int {\cal D}\varphi_{r,\vix}{\cal D}\varphi_{g,\vix}
\exp{\left(-S_{\rm H}-S_{\rm G}-S_{\rm F}\right)}; 
\label{Z_EPL}
\\
S_{\rm G}&=&-\kappa\sum_{\vix}\sum_{i =1}^3
\left( {\rm Tr}[U_{\vix}]{\rm Tr}[U_{\vix+ \vii}^\dagger]
+{\rm Tr}[U_{\vix}^\dagger]{\rm Tr}[U_{\vix+ \vii}]
\right), 
\nonumber\\
\label{SG_EPL}
\\
S_{\rm F}&=& \sum_{\vix} {\cal L}_{\rm F}({\bf x}), 
\label{SF_EPL}
\end{eqnarray}
%%%%%%%%%%%%%
where $U_\vix$ is the Polyakov line (loop) holonomy, the symbol $\mathbf{i}$ is an unit vector for $i$-th direction and ${\cal L}_{\rm F}$ is the fermionic Lagrangian density the concrete form of which will be shown later. 
The site index ${\bf  x}$ runs over a 3-dimensional lattice. 
Large (small) $\kappa$ in EPLM corresponds to high (low)
temperature~\cite{DeGrand} in QCD. 
Although the relation between $\kappa$ and $T$ is not simple, 
we regard the case with $\kappa \to 0$ and $M/T\to \infty$ as the zero-temperature limit in this paper. 

The Polyakov line holonomy $U_\vix$ is parameterized as ~\cite{Greensite:2014cxa}
%%%%%%%%%%%%%%%%
\begin{eqnarray}
U_\vix &=&{\rm diag} \left( e^{i\varphi_{r,\vix}},e^{i\varphi_{g,\vix}},e^{i\varphi_{b,\vix}}\right), 
\label{U_x}
\\
U_\vix^\dagger &=&{\rm diag}\left( e^{-i\varphi_{r,\vix}},e^{-i\varphi_{g,\vix}},e^{-i\varphi_{b,\vix}}\right), 
\label{U_x_dagger}
\end{eqnarray}
%%%%%%%%%%%%%%
with the condition $\varphi_{r,\vix}+\varphi_{g,\vix}+\varphi_{b,\vix}=0$. 
Instead of $U_\vix$ and $U_\vix^\dagger$, here the phase variables $\varphi_{r,\vix}$ and $\varphi_{g,\vix}$ are treated as dynamical variables.  
The Haar measure part ${\cal L}_{\rm H}$ is given by~\cite{Greensite:2014cxa}
%%%%%%%%%%%%%%%
\begin{eqnarray}
&&S_{\rm H}=\sum_{\vix} {\cal L}_{\rm H}({\bf x})
\label{HaarS} , \\
&&{\cal L}_{\rm H}({\bf x})=-\log{}\Bigl\{\sin^2{\left({\varphi_{r,\vix}-\varphi_{g,\vix}\over{2}}\right)}
\nonumber\\
&&\times \sin^2{\left({2\varphi_{r,\vix}+\varphi_{g,\vix}\over{2}}\right)}
\sin^2{\left({\varphi_{r,\vix}+2\varphi_{g,\vix}\over{2}}\right)
\Bigr\}}.
\label{HaarL}
\end{eqnarray}
%%%%%%%%%%%%%%
The (traced) Polyakov line (loop) $P_\vix$ and its conjugate $P_\vix^*$ are defined as 
%%%%%%%%%%%%%%
\begin{eqnarray}
P_\vix ={1\over{3}}{\rm Tr}\left[ U_\vix \right]&=&{1\over{3}}\left( e^{i\varphi_{r,\vix}}+e^{i\varphi_{g,\vix}}+e^{i\varphi_{b,\vix}}\right), 
\label{P_x}
\\
P_\vix^* ={1\over{3}}{\rm Tr}\left[ U_\vix^\dagger \right]&=&{1\over{3}}\left( e^{-i\varphi_{r,\vix}}+e^{-i\varphi_{g,\vix}}+e^{-i\varphi_{b,\vix}}\right). 
\nonumber\\
\label{P_x_conj}
\end{eqnarray}
%%%%%%%%%%%%%%%
For the fermionic Lagrangian density with the flavor dependent quark chemical potential $\mu_f$ and temperature $T$, 
we consider a logarithmic one of Ref. \cite{Greensite:2014cxa, Hirakida:2017bye}: 
%%%%%%%%%%%%%%
\begin{eqnarray}
{\cal L}_{\rm F}
&=&-2\sum_{f=u,d,s}\sum_{c=r,g,b} \Big\{\log{\Bigl(1+e^{\beta(\mu_f -M_f)+i\varphi_{c,\vix}}\Big)}
\nonumber\\
&&+\log{\Bigl(1+e^{-\beta(\mu_f +M_f)-i\varphi_{c,\vix}}\Big)}\Big\}
\nonumber\\
&=&-2\sum_{f=u,d,s}\Bigl\{\log{\Bigl(1+3e^{\beta(\mu_f -M_f)}P_\vix }
\nonumber\\
&&+3e^{2\beta(\mu_f -M_f)}P_\vix^* +e^{3\beta(\mu_f -M_f)}\Bigr)
\nonumber\\
&&+\log{\Bigl(1+3e^{-\beta(\mu_f +M_f)}P_\vix^*}
\nonumber\\
&&
+3e^{-2\beta(\mu_f +M_f)}P_\vix +e^{-3\beta(\mu_f+M_f)}\Bigr)\Bigr\}, 
\label{log_L}   
\end{eqnarray}
%%%%%%%%%%%%%%
where $M_f$ is the quark mass.     
In the limit $M_f\to \infty$, ${\cal L}_{\rm F}$ becomes real at $\mu =M_f$, since 
$e^{\beta (\mu-M_f)}=e^{2\beta (\mu-M_f)}=1$ and $e^{-\beta (\mu+M)}=e^{-2\beta (\mu+M_f)}=0$. 
The reality of ${\cal L_{\rm F}}$ at $\mu =M_f$ is related to the particle-hole symmetry~\cite{RF_PHS,Hirakida:2017bye}.  
In the zero-temperature limit, the antiparticle part of (\ref{log_L}) vanishes. 

It should be also noted that, at $\mu =M_f$, there is no dependence on $M_f/T$ in EPLM with symmetric flavors, if the antiquark contributions can be neglected.  
Hence, the point $\mu =M$ is the fixed point where physical quantities are not changed  when $M_f/T$ varies. 
Breaking of the flavor symmetry also breaks this invariance. 

The (dimensionless) quark-number density $n_q$ is obtained by 
%%%%%%%%%%%%%%%%%%%%%%%%%%%%%
\begin{eqnarray}
n_q &=&{1\over{\beta N_{\rm s}}}{\partial (\log{Z})\over{\partial \mu}}, 
\label{n_density}
\end{eqnarray}
%%%%%%%%%%%%%%%%%%%%%%%%%%%%%
where $N_{\rm s}$ is the number of the lattice spatial sites. 

As in the case of FHQM, $Z_3$-symmetrization (\ref{Z3symmetrization}) can be done for EPLM with three symmetric flavors.  
In this paper, we call the $Z_3$-symmetric EPLM "$Z_3$-EPLM". 
The fermionic Lagrangian density of $Z_3$ EPLM is given by
%%%%%%%%%%%%%%
\begin{eqnarray}
{\cal L}_{\rm F}
&=&-2\sum_{c=r,g,b} \Big\{\log{\Bigl(1+e^{\beta(3\mu -3M)+i3\varphi_{c,\vix}}\Big)}
\nonumber\\
&&+\log{\Bigl(1+e^{-\beta(3\mu +3M)-i3\varphi_{c,\vix}}\Big)}\Big\}. 
\label{log_L_Z3}   
\end{eqnarray}
%%%%%%%%%%%%%%
In the case of EPLM, the internal dynamical variables $\varphi_{c,\vix}$ are also threefold. 
Hence, this breaks the equivalence between $Z_3$-EPLM and the ordinary EPLM with threefold mass and threefold chemical potential.

%%%%%%%%%%%%%%%%%%%%%%%%%%%%%%%%%%%%%%%%%%%%%%%%%%%%%%%%%%%%%%%%%%%%%%%%%%%%%%%
\subsection{EPLM at $\kappa =0$}
\label{SEMI}
%%%%%%%%%%%%%%%%%%%%%%%%%%%%%%%%%%%%%%%%%%%%%%%%%%%%%%%%%%%%%%%%%%%%%%%%%%%%%%%

For $\kappa =0$, the partition function becomes simple, since the $S_{\rm G}$ term vanishes. 
For large $N_{\rm s}$ in which the periodic boundary condition is negligible, 
the integration over $\varphi_{r,\vix}$ and $\varphi_{g,\vix}$ can be performed 
independently for each site ${\bf x}$.     
The partition function turns out to be    
%%%%%%%%%%%%%%%%
\begin{eqnarray}
Z&=&\prod_{\vix} \left[\int_{-\pi}^\pi d\varphi_{r,\vix} \int_{-\pi}^\pi d\varphi_{g,\vix}
e^{-{\cal L}(\varphi_{r,\vix},  \varphi_{g \vix})}\right]
\nonumber\\
&=&\left[\int_{-\pi}^\pi du\int_{-\pi}^\pi dve^{-{\cal L}(u,v)}\right]^{N_{\rm s}}=z^{N_{\rm s}}, 
\label{Z_sa}
\end{eqnarray}
%%%%%%%%%%%%%%%%
where ${\cal L}={\cal L}_{\rm H}+{\cal L}_{\rm F}$, $N_{\rm s}$ is the number of the spatial lattice sites, and $z$ is the local partition function at one lattice site. 
It is known that the integral such as (\ref{Z_sa}) can be evaluated analytically~\cite{RF_PHS}. 
Hence, we call the equation "analytical" in this paper. 
However, the integral form is useful here, since we are interested in the mechanism of the sign problem.    

When ${\cal L}$ is not real,  instead of ${\cal L}$, we may use an approximate real Lagrangian ${\cal L}^\prime$ for constructing the probability density function.  
Then, the approximate partition function reads  
%%%%%%%%%%%%%%%%
\begin{eqnarray}
Z^\prime &=&\left[\int_{-\pi}^\pi du\int_{-\pi}^\pi dve^{-{\cal L}^\prime (u,v)}\right]^{N_{\rm s}}={z^\prime}^{N_{\rm s}} , 
\label{Z_sa_d}
\end{eqnarray}
%%%%%%%%%%%%%%%%
and the reweighting factor is $W=Z/Z^\prime=(z/z^\prime)^{N_{\rm s}}$.   
When we put ${\cal L}^\prime ={\rm Re}({\cal L})$, we obtain the reweighting (phase)  factor in the phase quenched (PQ) approximation.  
In this paper, we call the reweighting method with PQ approximation "PQRW".  
We examine how PQRW works in EPLM. 
For the brief review of PQRW, see appendix \ref{Aadd1}. 

Using Eq. (\ref{Z_sa}), the pressure $P$, the quark number density $n_q$, the scalar density $n_{\rm s}$, the averaged value 
of the Polyakov line $P_\vix$ and its conjugate $P_\vix$ are given by    
%%%%%%%%%%%%%%%%
\begin{eqnarray}
{P} ={T\over{N_{\rm s}}}\log{Z}=T\log{z}, 
\label{P_sa}
\end{eqnarray}
%%%%%%%%%%%%%%%%
%%%%%%%%%%%%%%%%
\begin{eqnarray}
n_q
={\int_{-\pi}^\pi du\int_{-\pi}^\pi dv\left(-T{\partial L\over{\partial \mu}}\right) e^{-{\cal L}(u,v)}\over{\int_{-\pi}^\pi du\int_{-\pi}^\pi dve^{-{\cal L}(u,v)}}}, 
\label{n_sa}
\end{eqnarray}
%%%%%%%%%%%%%%%%
%%%%%%%%%%%%%%%%
\begin{eqnarray}
n_{\rm s}
=\sum_{f=u,d,s}{\int_{-\pi}^\pi du\int_{-\pi}^\pi dv\left(T{\partial L\over{\partial M_f}}\right) e^{-{\cal L}(u,v)}\over{\int_{-\pi}^\pi du\int_{-\pi}^\pi dve^{-{\cal L}(u,v)}}}, 
\label{ns_sa}
\end{eqnarray}
%%%%%%%%%%%%%%%%
%%%%%%%%%%%%%%%%
\begin{eqnarray}
\langle P_\vix \rangle 
&=&{\int_{-\pi}^\pi du\int_{-\pi}^\pi dv P_\vix (u,v) e^{-{\cal L}(u,v)}\over{\int_{-\pi}^\pi du\int_{-\pi}^\pi dve^{-{\cal L}(u,v)}}}, 
\label{Phi_sa}
\end{eqnarray}
%%%%%%%%%%%%%%%%
%%%%%%%%%%%%%%%%
\begin{eqnarray}
\langle P_\vix^* \rangle 
&=&{\int_{-\pi}^\pi du\int_{-\pi}^\pi dv P_\vix^* (u,v) e^{-{\cal L}(u,v)}\over{\int_{-\pi}^\pi du\int_{-\pi}^\pi dve^{-{\cal L}(u,v)}}}, 
\label{Phi_sa_conj}
\end{eqnarray}
%%%%%%%%%%%%%%%%
respectively. 
These physical quantities are independent of $N_{\rm s}$, although $Z$ depends on $N_{\rm s}$. 
The analytical form of these quantities are modified when we consider the boundary condition. 
See Appendix~\ref{A1} for detail.  
 (It is easily seen that the modified results coincides with the equations above in the thermodynamical limit $N_{\rm s}\to \infty$. )

\begin{figure}[h]
\centering
\centerline{\includegraphics[width=0.40\textwidth]{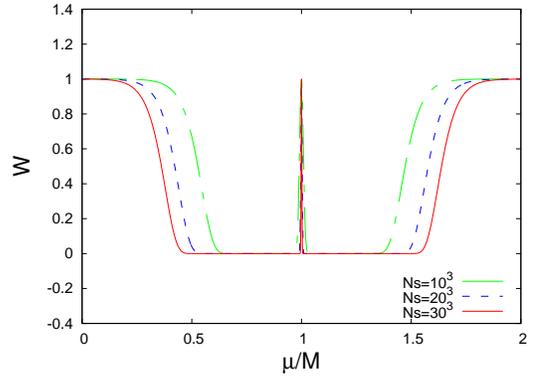}}
%\bigskip
%\bigskip
\caption{The $\mu$-dependence of the phase factor $W$. 
We set $M/T=10$. 
The dashed-dotted, dashed and solid lines represent the results with $N_{\rm s}=10^3,~20^3,30^3$, respectively. 
 }
 \label{EPLM_W_Ns}
\end{figure}

In Ref.~\cite{Hirakida:2017bye}, it was found that the phase factor and physical quantities at small $\kappa$ is very close to the ones at $\kappa =0$. 
Hence, we use Eqs.~(\ref{Z_sa})$\sim$(\ref{Phi_sa_conj}) as a phenomenological model for QCD with heavy quarks at low temperature. 
By use of these equations, we can discuss the sign problem from the results which are free from the sign problem.  
In the numerical calculations, we put $M_u=M_d=M_s=M$ unless otherwise mentioned. 
Figure \ref{EPLM_W_Ns} shows the $\mu$-dependence of the phase factor  $W$ in PQRW. 
Since $W$ depends on $N_{\rm s}$, we set $N_{\rm s}=10^3, 20^3$ and $30^3$.  
Around $\mu=M$, $W$ is small and the sign problem is serious in that region.   
However, due to the particle-hole symmetry, $S_{\rm F}$ is real and $W=1$ is always hold at $\mu =M$. 
The small $W$ indicates that the sign problem is serious when PQRW is used in simulations. 
When $N_{\rm s}$ increases, the sign problem be more serious. 
However, the $N_{\rm s}$-dependence becomes small when $N_s$ is large.  
Hence, we set $N_{\rm s}=30^3$ hereafter.   

%%%%%%%%%%%%%%%
%%%%%%%%%%%%%%%%

\begin{figure}[h]
\centering
\centerline{\includegraphics[width=0.40\textwidth]{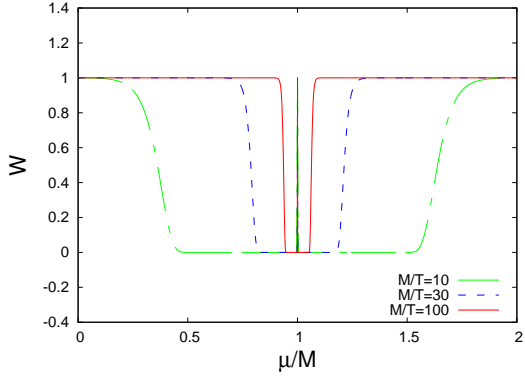}}
%\bigskip
%\bigskip
\caption{The $\mu$-dependence of the phase factor $W$.  
The dashed-dotted, dashed and solid lines represent the results with $M/T=10,30,100$, respectively. 
We set $N_{\rm s}=30^3$. 
 }
 \label{EPLM_W_M}
\end{figure}

Figures \ref{EPLM_W_M} shows the similar as Fig.~\ref{EPLM_W_Ns} but the one with fixed $N_s$. 
Roughly speaking, the sign problem is serious when $|\mu -M| < 5T$. 
Hence, when $M/T$ increases, the sign problem be weaker for fixed $\mu/M$. 
(However, the situation may be different when we compare them for fixed $\mu$. 
The phase factor $W$ can be larger for lighter quark than for heavier one when we compare them at fixed $\mu$. )

%%%%%%%%%%%%%%%%
%%%%%%%%%%%%%%%

\begin{figure}[h]
\centering
\centerline{\includegraphics[width=0.40\textwidth]{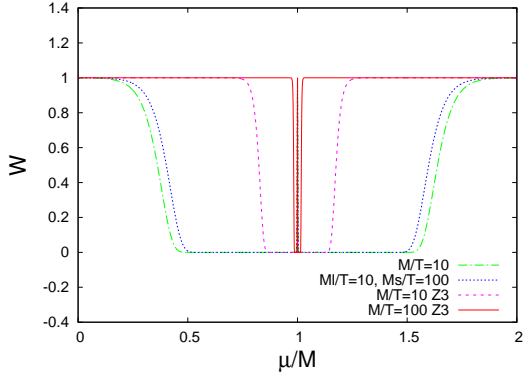}}
%\bigskip
%\bigskip
\caption{The $\mu$-dependence of the phase factor $W$.  
The dashed-dotted line represents the results with $M/T=10$ in EPLM , while the dotted line represents the result in EPLM with $M_u=M_d=M_l=M=10T$ and $M_s=10M$. 
The dashed and solid lines represent the result in $Z_3$-EPLM with  $M/T=10, 100$, respectively. 
We set $N_{\rm s}=30^3$. 
 }
 \label{EPLM_W_model}
\end{figure}

Figures \ref{EPLM_W_model} shows the result in EPLM without three flavor symmetry.  
When the strange quark mass $M_s$ is larger than the light quark mass $M_l$, the sign problem becomes weaker, but the change is not so large. 
This indicates that the lighter quark dominates the sign problem.  
In Fig.~\ref{EPLM_W_model}, the results in $Z_3$-EPLM are also shown.  
It can be seen that the $Z_3$-symmetrization makes the sign problem weak drastically.  
In the case of $Z_3$-EPLM with $M/T=100$, the sign problem almost vanishes except for the vicinity of $\mu =M$. 
In this case, from $W$ itself, we see that nonanalyticity certainly happens just at $\mu =M$. 

%%%%%%%%%%%%%%%%
%%%%%%%%%%%%%%%

Comparing Fig. \ref{EPLM_W_model} with Fig. \ref{EPLM_W_M}, we see that the sign problem is somewhat weaker in $Z_3$-EPLM with $M/T=10$ than EPLM with $M/T=30$. 
This is because, different from the number density, the partition function in $Z_3$-EPLM is close to the one in EPLM with threefold mass and three fold chemical potential but one flavor. 
As well as the decrease of $N_{\rm s}$, the decreases of $N_f$ make the sign problem weaker.  

\begin{figure}[h]
 \centering
 \includegraphics[width=0.40\textwidth]{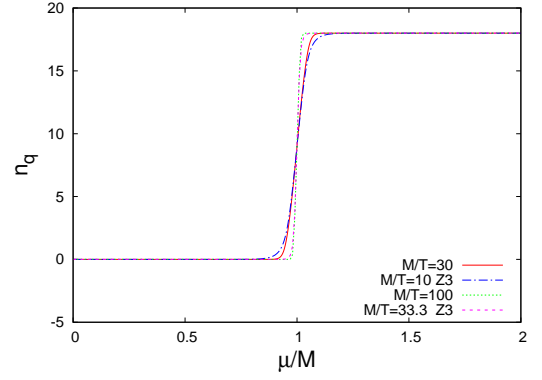}
%\vspace{10mm}
 \caption{The $\mu$-dependence of the quark number density $n_{\rm q}$.  
 The solid and dotted lines represent the results in EPLM with $M/T=30$ and 100,   respectively, while  the dash-dotted and dashed lines represent the results in $Z_3$-EPLM with $M/T=10$ and 33.3,  respectively. 
 }
 \label{EPLM_rho}
\end{figure}

Figures~\ref{EPLM_rho} shows the $\mu$-dependence of the quark number density $n_q$. 
It is seen that $n_q$ abruptly increases at $\mu =M$, when $M/T$ is large.  
The results are antisymmetric with respect the point $\mu =M$. 
This property is the consequence of the particle-hole symmetry. 
Due to the effect of the gauge field $\varphi_{c,\vix}$, the equivalence between $Z_3$-EPLM and EPLM with threefold mass and threefold chemical potential is slightly broken, but the difference be smaller when $M/T$ becomes larger.  

%%%%%%%%%%%%%%%
%%%%%%%%%%%%%%%

\begin{figure}[h]
\centering
\centerline{\includegraphics[width=0.40\textwidth]{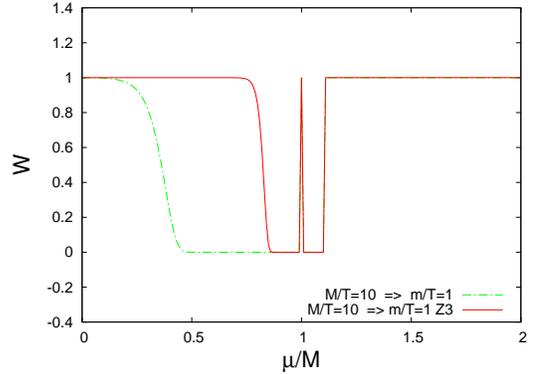}}
%\bigskip
%\bigskip
\caption{The $\mu$-dependence of the phase factor $W$.  
At $\mu/M=1.1$, the quark mass changes from $M (=10T)$ from $m (=T)$.   
The dashed-dotted and solid lines represent the results in EPLM  and $Z_3$-EPLM, respectively. 
We set $N_{\rm s}=30^3$. 
 }
 \label{EPLM_W_chiral}
\end{figure}

In the heavy quark model, the scalar density $n_{\rm s}$ is almost the same as $n_{\rm q}$, since the effects of the spatial momentum and the vacuum fluctuations are absent and the antiquark contribution is negligible.  
If it couples to the quark field, it can make the quark mass smaller. 
In Fig.~\ref{EPLM_W_chiral}, the results in EPLM and $Z_3$-EPLM in which quark mass changes from large mass $M (=10T)$ to small one $m (=T)$ at $\mu /M=1.1$.  
Remember that $W$ can be larger for lighter quark than the one for heavier quark when we compare them at fixed $\mu$. 
Due to the change of the quark mass,  the symmetry with respect to the line $\mu /M =1$ is broken.  
In EPLM and $Z_3$-EPLM, the breaking of symmetry around $\mu =M$ may indicate a nontrivial change of the system. 
However, in QCD, such symmetry is not expected at the beginning. 
Hence, we should control the trivial sign problem caused by the formation of Fermi sphere anyway. 

%%%%%%%%%%%%%%%%
%%%%%%%%%%%%%%%

\begin{figure}[h]
 \centering
\includegraphics[width=0.40\textwidth]{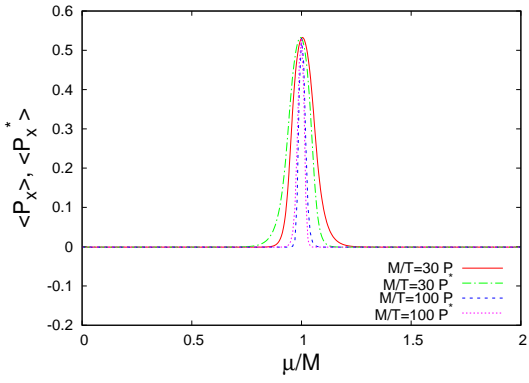}
%\vspace{8mm}
 \caption{The $\mu$-dependence of the averaged values $\langle P_x\rangle$ and  $\langle P_x^*\rangle$, when EPLM. is used.  
 The solid and dashed lines represent $\langle P_x\rangle$ with $M/T=30, 100$, respectively. 
The dash-dotted and dotted lines represent $\langle P_x^*\rangle$ with $M/T=30, 100$, respectively. 
 }
 \label{EPLM_Pol_Polc}
\end{figure}

Figures \ref{EPLM_Pol_Polc} shows the $\mu$-dependence of the averaged value $\langle P_{\vix} \rangle$ and $\langle P_{\vix}^* \rangle$ in EPLM. 
It is seen that the both quantities somewhat large in the vicinity of $\mu =M$. 
$\langle P_{\vix} \rangle$ has a maximum in the region $\mu >M$, while $\langle P_{\vix}^* \rangle$ does in the region $\mu <M$.  
This property is  also a consequence of the particle-hole symmetry. 
In $Z_3$-EPLM, the expectation value of the Polyakov-loop vanishes due to the exact $Z_3$-symmetry.

%%%%%%%%%%%%%%
%%%%%%%%%%%%%%%

When all $\varphi_{c,\vix}$ vanish, $\langle P_\vix\rangle $ and $\langle P_\vix^* \rangle $ becomes 1.  
The case corresponds to the ordered phase and the sign problem does not happen since all $\varphi_{c.\vix}$ vanish. 
However, the absolute values of the Polyakov line is far from 1 and the $\varphi_{c, \vix}$ fluctuates almost randomly. 
This causes the serious sign problem around $\mu =M$.  

In Fig.~\ref{EPLM_Pol_Polc}, when $M/T$ increases, 
$\langle P_\vix\rangle $ and $\langle P_\vix^* \rangle $ becomes smaller at $\mu \neq M$, but they do not change at $\mu =M$ since the point $\mu =M$ is the fixed point. 
Hence, in the limit $M/T\to \infty$, the Polyakov line (and its conjugate) at $\mu \neq M$ and all the other quantities at any $\mu$ approach to the ones in $Z_3$-EPLM,  however, the Polyakov line (and its conjugate) at $\mu =M$ has different value in two models.    
It seems that the Polyakov line on the nonanalytical point can detect the difference of the boundary condition even in the  zero-temperature limit. 

\begin{figure}[h]
\centering
\includegraphics[width=0.40\textwidth]{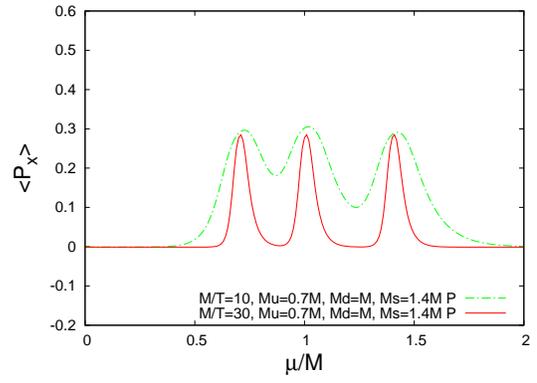}
%\vspace{8mm}
\caption{The $\mu$-dependence of the averaged values $\langle P_x\rangle$ in the nonsymmetric flavor EPLM with mass $M_u=0.7M$, $M_d=M$, $M_s=1.4M$. 
The dash-dotted and solid line represent the results with $M/T=10,30$, respectively. }
\label{EPLM_Pol_Polc_Nf111}
\end{figure}

It should be remarked that the existence of the fixed point at $\mu =M$ has an important role in the anomalous phenomena mentioned above. 
When the flavor symmetry is broken, the exact fixed point disappears. 
Figure~\ref{EPLM_Pol_Polc_Nf111} shows the $\mu$-dependence of $\langle P_\vix\rangle $ in EPLM with nonsymmetric flavors. 
In this case, the absolute values of these quantities are smaller than those in the symmetric flavor case. 
Furthermore, since $\mu =M_f~(f=u,d,s)$ is an approximate fixed point but not an exact one, the maximum values of these quantities decrease as $M_f/T$ increases. 
Hence, if we take the effect of the spatial momentum of quarks into account and take the zero temperature limit, the expectation value of the Polyakov line may vanish. 
(Althouh we do not show the result, the $\mu$-dependence of  $\langle P_\vix^*\rangle $ shows the similar tendency. )

%%%%%%%%%%%%%%
%%%%%%%%%%%%%%%

%%%%%%%%%%%%%%%%%%%%%%%%%%%%%%%%%%%%%%%%%%%%%%%%%%%%%%%%%%%%%%%%
\subsection{Relation between sign problem and nonanalyticity at zero temperature}
\label{RBSN}
%%%%%%%%%%%%%%%%%%%%%%%%%%%%%%%%%%%%%%%%%%%%%%%%%%%%%%%%%%%%%%%%

\begin{figure}[h]
 \centering
\includegraphics[width=0.40\textwidth]{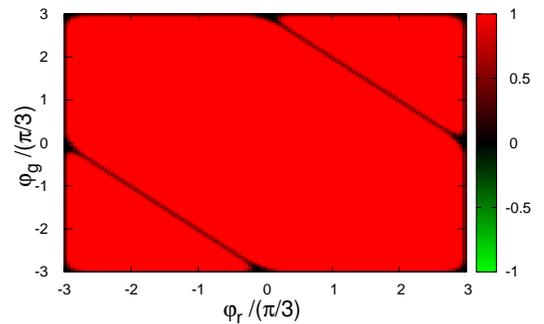}
%\vspace{10mm}
 \caption{The $\varphi_r$ and $\varphi_g$-dependence of $e^{-\cal{L}_{\rm F}}/(|e^{-\cal{L}_{\rm F}}|+\epsilon )$ in EPLM. 
 We set $M/T=100$ and $\mu =M$. 
 }
 \label{action_EPLM_M}
\end{figure}

The local partition function $z$ can be written as   
%%%%%%%%%%%%%%%%
\begin{eqnarray}
z=\int_{-\pi}^\pi d\varphi_r \int_{-\pi}^{\pi} d\varphi_g  e^{-{\cal L}_{\rm H}}e^{-{\cal L}_{\rm F}}, 
\label{localz}
\end{eqnarray}
%%%%%%%%%%%%%%%%
where the spatial index $\vix$ is ommited for simplicity of the notation. 
The factor $f=e^{-{\cal L}_{\rm F}}$ is given by
%%%%%%%%%%%%%%%%
\begin{eqnarray}
f(\varphi_r,\varphi_g)&=&\exp{\left(2\sum_{f,c}\log{(1+e^{(\mu_{f,c}-M)/T})}\right)}, 
\label{microz}
\end{eqnarray}
%%%%%%%%%%%%%%%%
where $\mu_{f,c}=\mu+i\varphi_{c}T$ in EPLM and $\mu_{f,c}=\mu+i(\theta_f+i\varphi_{c})T$ in $Z_3$-EPLM. 
This factor can be used as a microscopic probability density function in numerical simulation if the sign problem is absent.  
Figure~\ref{action_EPLM_M} shows $f/(|f|+\epsilon )$ at $\mu =M$ in EPLM with $M/T=100$, where $\epsilon$ is a positive infinitesimal constant.   
Note that, due to the particle-hole symmetry, $f$ is real and is nonnegative at $\mu =M$.  
Hence $f/(|f|+\epsilon )$ is 1 unless $f=0$. 
The set of zeros forms a line structure.  
If one of ${\rm Im}(\mu_{f,c})$ is equal to $(2k+1)\pi$, where $k$ is an integer, $f$ becomes zero at $\mu =M$. 
This condition corresponds to the horizontal and vertical black lines at the edges in the figure.   
Furthermore, $f$ is also zero when $\varphi_g=(2k+1)\pi -\varphi_r$ is satisfied. 
This condition corresponds to the black oblique lines in the figure.  

%%%%%%%%%%%%%%%
%%%%%%%%%%%%%%%

\begin{figure}[h]
 \centering
\includegraphics[width=0.40\textwidth]{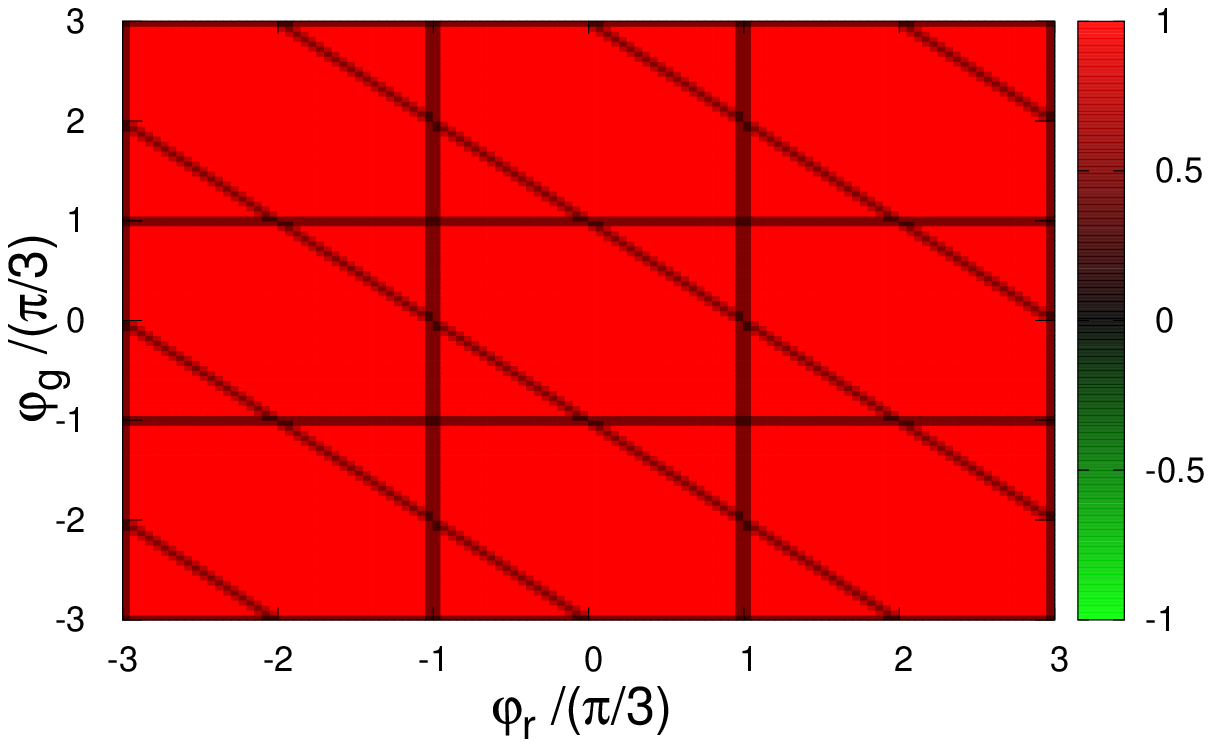}
%\vspace{10mm}
 \caption{The $\varphi_r$ and $\varphi_g$-dependence of $e^{-\cal{L}_{\rm F}}/(|e^{-\cal{L}_{\rm F}}|+\epsilon )$ in $Z_3$-EPLM. 
 We set $M/T=33.3$ and $\mu =M$. 
 }
 \label{action_EPLM_M_Z3}
\end{figure}

Figure~\ref{action_EPLM_M_Z3} shows the same  as Fig.~\ref{action_EPLM_M} but $Z_3$-EPLM with $M/T=33.3$ is used.  
Due to the $Z_3$-symmetry, the zero-structure in the region where $\varphi_g={2k-1\over{3}}\pi\sim{2k+1\over{3}}\pi$ and ${2l-1\over{3}}\pi\sim {2l+1\over{3}}\pi$ with $k,l=-1,0,1$ is similar to the one of the EPLM result in the region where $\varphi_r=-\pi\sim\pi$ and $\varphi_g=-\pi\sim\pi$.  

%%%%%%%%%%%%%%%
%%%%%%%%%%%%%%%

The structure of zeros of $f$ in Fig.~\ref{action_EPLM_M_Z3} is $Z_3$-symmetric but is not in Fig.~\ref{action_EPLM_M}. 
Hence, it can be said that the Polyako-line and its conjugate can detect the microscopic structure of zeros of $f$ at $\mu =M$ and decide to be or not to be finite in the limit of $M/T\to \infty$, although the other quantities are not sensible to the structure. 

\begin{figure}[h]
 \centering
\includegraphics[width=0.40\textwidth]{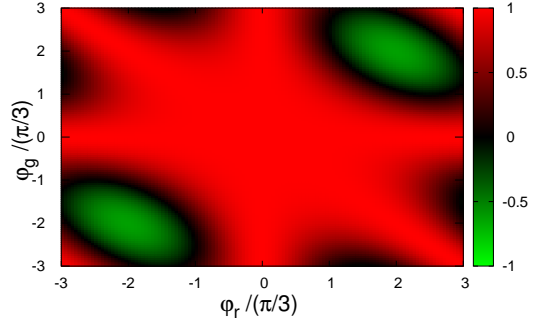}
%\vspace{10mm}
 \caption{The $\varphi_r$ and $\varphi_g$-dependence of ${\rm Re}(f)/(|f|+\epsilon)$ in EPLM. 
 We set $M/T=100$ and $\mu =0.98M$. 
 }
 \label{action_EPLM_098M}
\end{figure}

It should be noted that the zeros of $f$ at $\mu =M$ themselves do not induce the sign problem. 
However, in the vicinity of $\mu =M$, the situation is changed drastically.  
Suppose $f=0$ at $\varphi_r=\varphi_{r0}$ and $\varphi_{g0}$ when $\mu =M$. 
Then, the absolute value of ${\rm Im}[{\cal{L}_{\rm F}}(\varphi_{r,0},\varphi_{g,0})]$ may be still small at $\mu =M+\Delta \mu$ when $|\Delta \mu |$ is small enough.  
However, expanding ${\cal L}_{\rm F}(\varphi_r,\varphi_g)$ at $\mu =M+\Delta \mu$ in term of $\Delta \varphi_r=\varphi_r-\varphi_{r,0}$ and $\Delta \varphi_g=\varphi_g-\varphi_{g,0}$, we obtain 
%%%%%%%%%%%%%%%%
\begin{eqnarray}
&&{\cal L}_{\rm F}(\varphi_r,\varphi_g)={\cal L}(\varphi_{r,0},\varphi_{g,0})
\nonumber\\
&&+i{\cal L}_{\rm F}^\prime (\varphi_{r,0}\varphi_{g,0})(\Delta \varphi_r+\Delta \varphi_g)
\nonumber\\
&&-{1\over{2}}{\cal L}_{\rm F}^{\prime\prime} (\varphi_{r,0}\varphi_{g,0})(\Delta \varphi_{r,0}+\Delta \varphi_{g,0})^2
\nonumber\\
&&-i{1\over{6}}{\cal L}_{\rm F}^{\prime\prime\prime} (\varphi_{r,0}\varphi_{g,0})(\Delta \varphi_{r,0}+\Delta \varphi_{g,0})^3+\cdots, 
\label{expandL}
\end{eqnarray}
%%%%%%%%%%%%%%%%
where $\prime$ denotes the differentiation with respect to $\mu$. 
Since structure of the fermionic Lagrangian ${\cal L}_{\rm F}$ in EPLM is similar to the pressure of FHQM discussed in Sec.~\ref{FHF}, the odd coefficients in the expansion (\ref{expandL}) have divergent behavior and ${\rm Im}({\cal L}_{\rm F})$ can be large at $\mu =M+\Delta \mu$. 
This makes the sign problem serious in the vicinity of $\mu =M$. 
Figure~\ref{action_EPLM_098M} shows ${\rm Re}(f)/(|f|+\epsilon)$ at $\mu =0.98M$ in EPLM with $M/T=100$.  
There are area-like regions where ${\rm Re}(f)$ is negative. 
These regions  make $|z|$ smaller than the quenched one $z^\prime$ and induce a serious sign problem. 
(Note that $W$ will be almost zero in the large $N_{\rm s}$ limit, even if $|z|$ is slightly smaller than $z^\prime$. ) 

%%%%%%%%%%%%%%%
%%%%%%%%%%%%%%%

\begin{figure}[h]
 \centering
\includegraphics[width=0.40\textwidth]{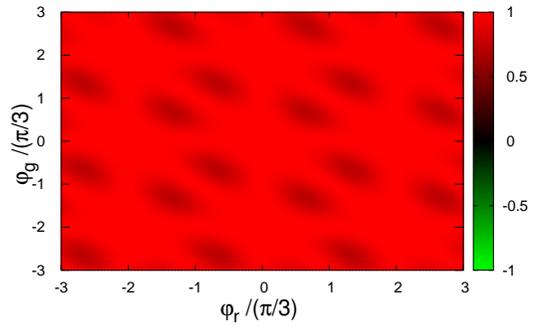}
%\vspace{10mm}
 \caption{The $\varphi_r$ and $\varphi_g$-dependence of ${\rm Re}(f)/(|f|+\epsilon)$ in $Z_3$-EPLM. 
 We set $M/T=33.3$ and $\mu =0.98M$. 
 }
 \label{action_EPLM_098M_Z3}
\end{figure}

Figure~\ref{action_EPLM_098M_Z3} shows the same  as Fig.~\ref{action_EPLM_098M} but $Z_3$-EPLM with $M/T=33.3$ is used.  
As is the same as in Fig.~\ref{action_EPLM_M_Z3},  $Z_3$-symmetric structure is seen also in this figure. 
The minimum value of ${\rm Re}(f)/(|f|+\epsilon) $ is much larger than that in Fig.~\ref{action_EPLM_098M} and is not negative. 
Hence, the sign problem is not so strong in this case. 
When we set $M/T=100$ in $Z_3$-EPLM,  ${\rm Re}(f)/|f| =1$ is realized anywhere, almost perfectly, the sign problem almost vanishes at $\mu =0.98M$. 

%%%%%%%%%%%%%%%
%%%%%%%%%%%%%%%

%%%%%%%%%%%%%%%%%%%%%%%%%%%%%%%%%%%%%%%%%%%%%%%%%%%%%%%%%%%%%%%%
\subsection{Effects of imaginary quark chemical potential}
\label{IQCP}
%%%%%%%%%%%%%%%%%%%%%%%%%%%%%%%%%%%%%%%%%%%%%%%%%%%%%%%%%%%%%%%%

\begin{figure}[h]
\centering
\centerline{\includegraphics[width=0.40\textwidth]{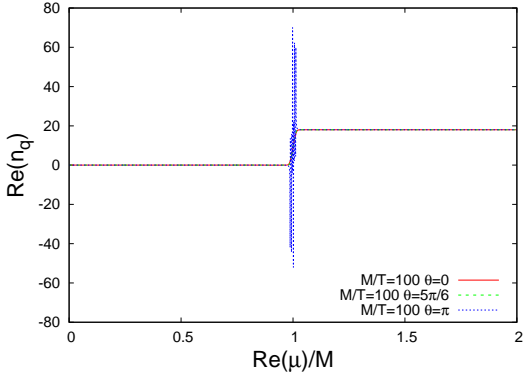}}
%\bigskip
%\bigskip
\caption{The $\mu$-dependence of ${\rm Re}(n_{\rm q})$ in EPLM with $M/T=100$.  
The solid, dashed and dotted lines represent the result with $\theta =0$, $5\pi/6$ and $\pi$, respectively.  
Three lines almost coincide each other except in the neighborhood of ${\rm Re}(\mu ) =M$. 
 }
\label{EPLM_Re_n}
\end{figure}

When the imaginary quark chemical potential $i\theta T$ is introduced, the reality of the grand canonical partition function $Z$ and physical quantities is not ensured in general.  
Figure~\ref{EPLM_Re_n} shows the complex chemical potential dependence of ${\rm Re}(n_q)$ in EPLM with $M/T=100$. 
Except in the neighbor of ${\rm Re} (\mu )=M$, the results for finite $\theta$ coincide with that for $\theta =0$. 
This is because the imaginary chemical potential, which is equivalent to the change of the boundary condition, is irrelevant in the zero temperature limit. 
In the neighborhood of ${\rm Re}(\mu)=M$ the result for $\theta =\pi$ vibrates violently. 
The maximum value of $|{\rm Re}(n_{\rm q})|$ is much larger than the degree of freedom of quarks, namely, $2N_fN_c=18$. 
In this figure, we show the result at $0.001$ intervals in the horizontal axis. 
The singular behavior for $\theta =\pi$ depends strongly on the interval we use. 
When the interval is smaller, the maximum value of $|{\rm Re}(n_{\rm q})|$ is larger. 
For $\theta =\pi$, large quark number density can be induced at the singular point.  
This phenomenon happens also at ${\rm Re}(\mu ) =-M$. 

%%%%%%%%%%%%%%%%
%%%%%%%%%%%%%%%

\begin{figure}[h]
\centering
\centerline{\includegraphics[width=0.40\textwidth]{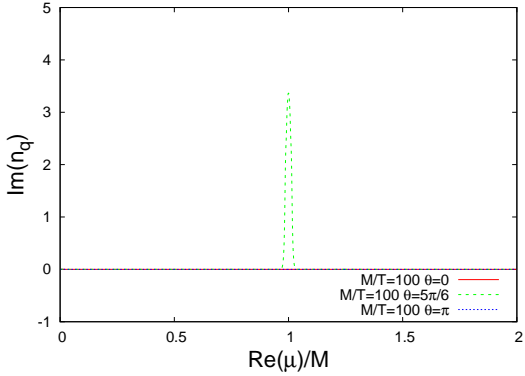}}
%\bigskip
%\bigskip
\caption{The $\mu$-dependence of ${\rm Im}(n_{\rm q})$ in EPLM wit $M/T=100$.  
The solid, dashed and dotted lines represent the result with $\theta =0$, $5\pi/6$ and $\pi$, respectively.  
 }
\label{EPLM_Im_n}
\end{figure}

Figure~\ref{EPLM_Im_n} shows the complex chemical potential dependence of ${\rm Im}(n_q)$ in EPLM with $M/T=100$. 
When $\theta =0$ or $\pi$, ${\rm Im}(n_{\rm q})=0$.  
For $\theta ={5\over{6}}\pi$, ${\rm Im}(n_q)$ is finite only in the vicinity of ${\rm Re}(\mu )=M$. 
This is also because the imaginary chemical potential which induces the imaginary part of  ${\rm Im}(n_q)$ is irrelevant in the zero temperature limit. 

%%%%%%%%%%%%%%%%
%%%%%%%%%%%%%%%

\begin{figure}[h]
\centering
\centerline{\includegraphics[width=0.40\textwidth]{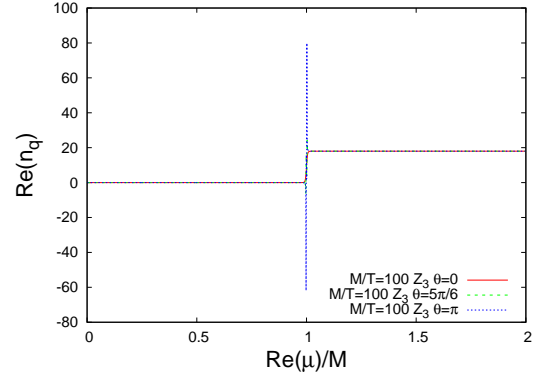}}
%\bigskip
%\bigskip
\caption{The $\mu$-dependence of ${\rm Re}(n_{\rm q})$ in $Z_3$-EPLM wit $M/T=100$.  
The solid, dashed and dotted lines represent the result with $\theta =0$, $5\pi/6$ and $\pi$, respectively.  
 }
\label{Z3EPLM_Re_n}
\end{figure}

\begin{figure}[h]
\centering
\centerline{\includegraphics[width=0.40\textwidth]{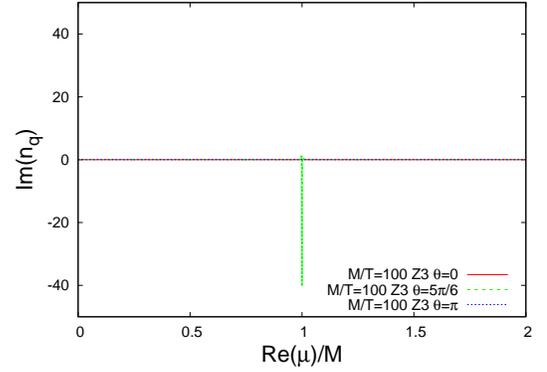}}
%\bigskip
%\bigskip
\caption{The $\mu$-dependence of ${\rm Im}(n_{\rm q})$ in $Z_3$-EPLM wit $M/T=100$.  
The solid, dashed and dotted lines represent the result with $\theta =0$, $5\pi/6$ and $\pi$, respectively.  
 }
\label{Z3EPLM_Im_n}
\end{figure}

Figures~\ref{Z3EPLM_Re_n} and \ref{Z3EPLM_Im_n} shows the $\mu$-dependence of the number density in $Z_3$-EPLM with $M/T=100$. 
The number density $n_{\rm q}$ has similar properties as the one in EPLM, but the singularity for $\theta =\pi$ is very sharp.   
We also observe the oscillating behavior in ${\rm Im}(n_{\rm q})$ for $\theta={5\over{6}}\pi$.  

%%%%%%%%%%%%%%%%
%%%%%%%%%%%%%%%

%%%%%%%%%%%%%%%%
%%%%%%%%%%%%%%%

\begin{figure}[h]
\centering
\centerline{\includegraphics[width=0.40\textwidth]{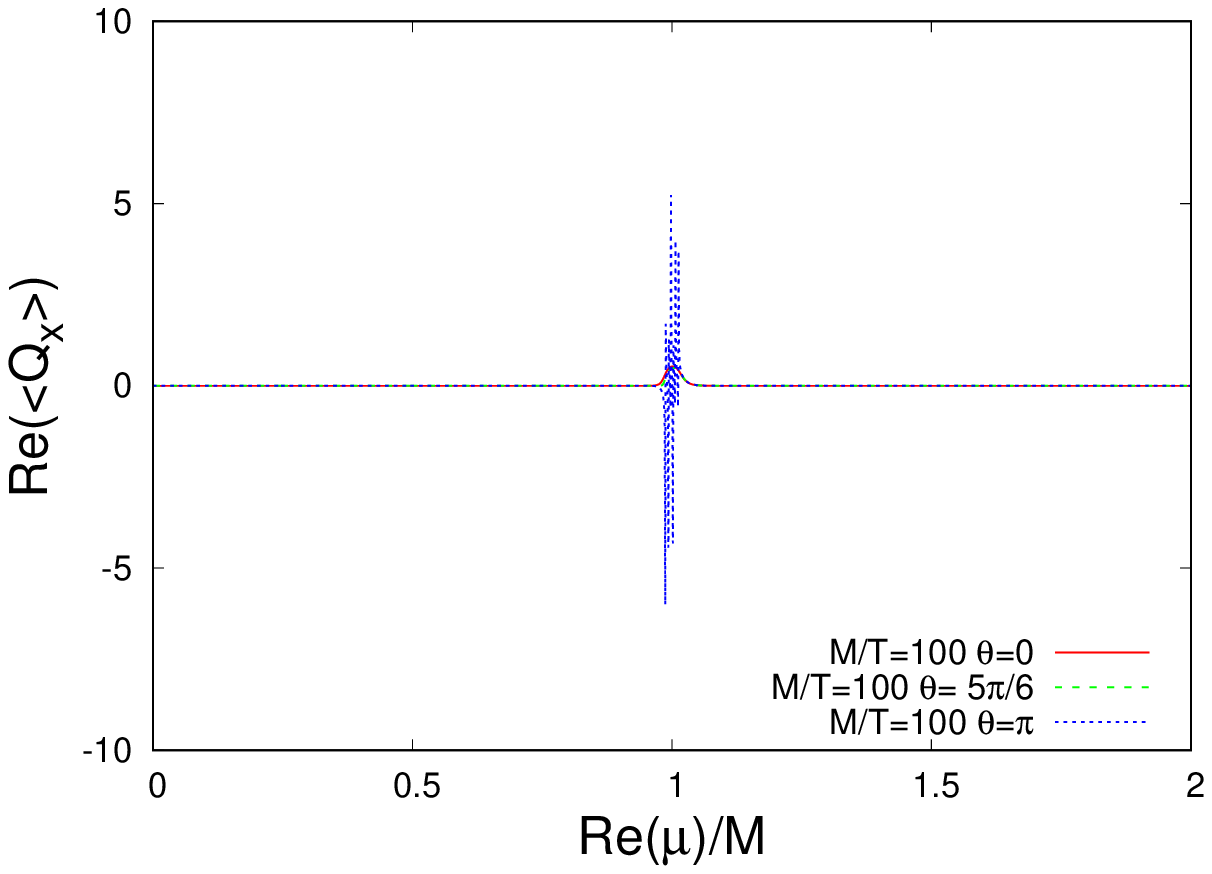}}
%\bigskip
%\bigskip
\caption{The $\mu$-dependence of ${\rm Re}(Q_\vix )$ in EPLM wit $M/T=100$.  
The solid, dashed and dotted lines represent the result with $\theta =0$, $5\pi/6$ and $\pi$, respectively.  
Three lines almost coincide each other except for the vicinity of ${\rm Re}(\mu ) =M$. 
 }
\label{EPLM_Re_mPol}
\end{figure}

\begin{figure}[h]
\centering
\centerline{\includegraphics[width=0.40\textwidth]{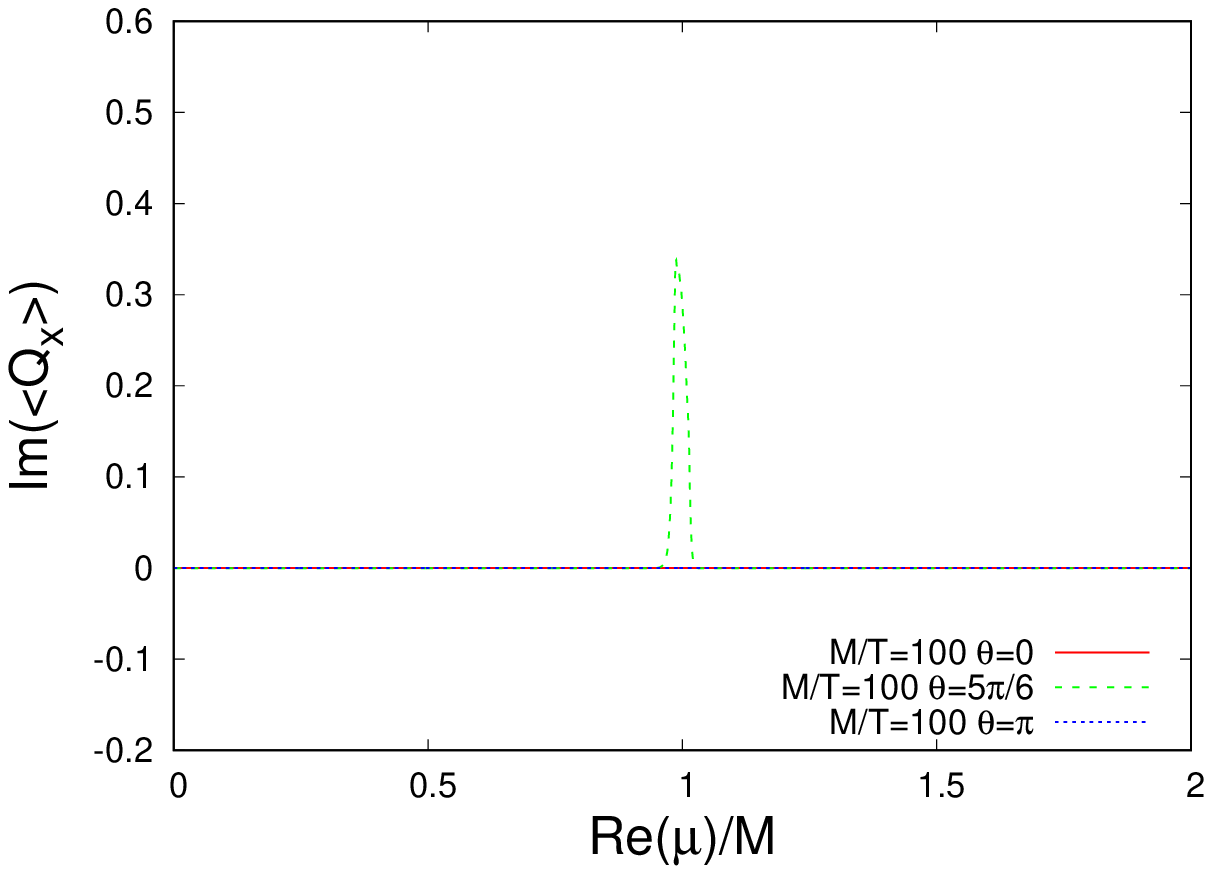}}
%\bigskip
%\bigskip
\caption{The $\mu$-dependence of ${\rm Im}(Q_\vix )$ in EPLM wit $M/T=100$.  
The solid, dashed and dotted lines represent the result with $\theta =0$, $5\pi/6$ and $\pi$, respectively.  
 }
\label{EPLM_Im_mPol}
\end{figure}

It is known that the introduction of the imaginary chemical potential rotates the Polyakov line in the complex plane. 
Hence, we use the modified Polyakov line $Q_\vix =e^{i\theta}P_\vix$ instead of $P_\vix$ itself.  
Note that $Q_\vix$ has the Roberge-Weiss (RW) periodicity, namely, $\langle Q_\vix (\theta +{2\over{3}}\pi)\rangle =\langle Q_x(\theta )\rangle$, but $P_\vix$ does not~\cite{Sakai:2008py,Sakai:2008um,Kouno:2009bm,Kashiwa:2019dqn}. 
 Figures~\ref{EPLM_Re_mPol} and~\ref{EPLM_Im_mPol} show the complex chemical potential dependence of ${\rm Re}(\langle Q_x\rangle )$ and ${\rm Im}(\langle Q_x\rangle )$ in EPLM with $M/T=100$. 
In these figures, the same tendency is seen as in the case of $n_{\rm q}$. 
In $Z_3$-EPLM, $\langle Q_x \rangle$ is always zero due to the exact $Z_3$-symmetry. 

%%%%%%%%%%%%%%%%
%%%%%%%%%%%%%%%

%%%%%%%%%%%%%%%%
%%%%%%%%%%%%%%%

\begin{figure}[h]
\centering
\centerline{\includegraphics[width=0.40\textwidth]{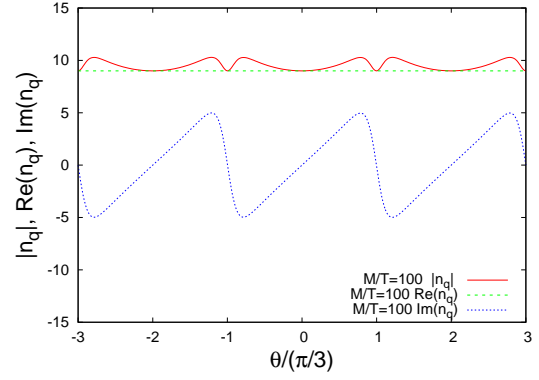}}
%\bigskip
%\bigskip
\caption{The $\theta$-dependence of $|n_{\rm q}|$ (solid line), ${\rm Re}(n_{\rm q})$ (dashed line) and ${\rm Im} (n_{\rm q})$ (dotted line) at ${\rm Re}(\mu )=M$ in EPLM. 
We set $M/T=100$. 
 }
\label{EPLM_period}
\end{figure}

\begin{figure}[h]
\centering
\centerline{\includegraphics[width=0.40\textwidth]{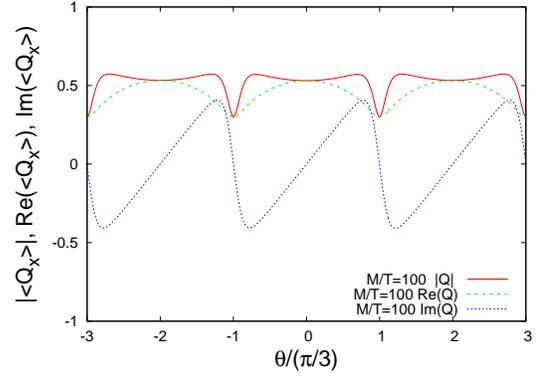}}
%\bigskip
%\bigskip
\caption{The $\theta$-dependence of 
$|\langle Q_\vix\rangle |$ (solid line), ${\rm Re}(\langle Q_\vix \rangle )$ (dashed line) and ${\rm Im} (\langle Q_\vix \rangle)$ (dotted line) at ${\rm Re}(\mu )=M$ in EPLM. 
We set $M/T=100$. 
 }
\label{EPLM_period_Q}
\end{figure}

Figures~\ref{EPLM_period} and~\ref{EPLM_period_Q} show the $\theta$-dependence of $n_{\rm q}$ and $\langle Q_x\rangle$ at $\mu =M$, respectively.  
The RW periodicity is clearly seen in these figures. 
Note that the imaginary parts of $n_{\rm q}$ and $Q_\vix$ are the indicators of the RW-transition~\cite{Sakai:2008py,Sakai:2008um,Kouno:2009bm,Kashiwa:2019dqn}. 
Here the RW periodicity is smooth and this property is not changed by varying $M/T$ since the point ${\rm Re}(\mu )=M$ is the fixed point. 
Hence, it is expected that the RW transition does not occur even in the limit $M/T\to \infty$.   
Of course, this may be natural since we have set $\kappa =0$. 
If the interaction between gauge field is switched on, a nontrivial transition may happen. 
The study in EPLM with finite $\theta$ and nonvanishing $\kappa$ at low temperature limit is an interesting problem in future. 
(The study on the $Z_N$-spin model with the external complex field and the interaction between the spins can be found in Ref.~\cite{deForcrand:2017rfp}. In that case, the hard sign problem induced by the external complex field was found. )

%%%%%%%%%%%%%%%%
%%%%%%%%%%%%%%%

%%%%%%%%%%%%%%%%
%%%%%%%%%%%%%%%

\begin{figure}[h]
\centering
\centerline{\includegraphics[width=0.40\textwidth]{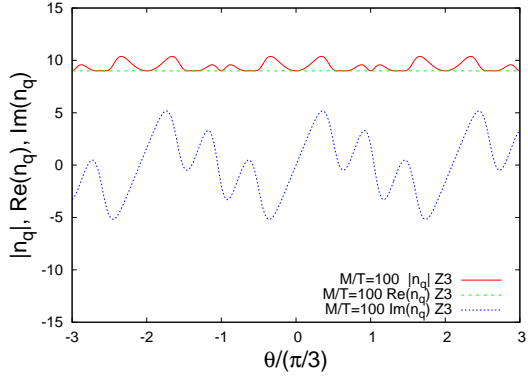}}
%\bigskip
%\bigskip
\caption{The $\theta$-dependence of $|n_{\rm q}|$ (solid line), ${\rm Re}(n_{\rm q})$ (dashed line) and ${\rm Im} (n_{\rm q})$  (dotted line) at ${\rm Re}(\mu )=M$ in $Z_3$-EPLM. 
We set $M/T=100$.   
 }
\label{Z3EPLM_period}
\end{figure}

Figure~\ref{Z3EPLM_period} shows the $\theta$-dependence of $n_{\rm q}$ at $\mu =M$ in $Z_3$-EPLM.  
The RW periodicity is seen in these figures. 
Furthermore, in ${\rm Im}(n_{\rm q})$, the higher frequency mode with the period ${2\over{9}}\pi$ is clearly seen. 
This property is related to the $Z_3$-symmetry. 
The $\theta$-dependence of ${\rm Im}(n_{\rm q})$ is very sensitive to the $Z_3$-symmetry structure at ${\rm Re}(\mu ) =M$.

%%%%%%%%%%%%%%%%
%%%%%%%%%%%%%%%

%%%%%%%%%%%%%%%%%%%%%%%%%%%%%%%%%%%%%%%%%%%%%%%%%%%%%%%%
%%%% summary
%%%%%%%%%%%%%%%%%%%%%%%%%%%%%%%%%%%%%%%%%%%%%%%%%%%%%%%%
\section{Summary}
\label{summary}
%%%%%%%%%%%%%%%%%%%%%%%%%%%%%%%%%%%%%%%%%%%%%%%%%%%%%%%%

In this paper, we have studied the non-analyticity and the sign problem in the $Z_3$-symmetric heavy quark model at low temperature and examined how the $Z_3$-symmetrized models approach to the original ones in the zero temperature limit.  
For the free fermion quark model (FHQM), the non-analyticity at $\mu =M$ is related to the existence of zeros of the grand canonical partition function $Z$ at finite temperature and complex chemical potential. 
By $Z_3$-symmetrization, the zeros are threefold, but the $Z_3$ symmetric FHQM ($Z_3$-FHQM) is equivalent to the original one with threefold quark mass and threefold quark chemical potential. 
Therefore, $Z_3$-FHQM naturally approaches to the original one in the zero temperature limit.  

We also examined the three flavor effective Polyakov-line model (EPLM ) with $\kappa =0$.  
In $Z_3$-symmetric EPLM ($Z_3$-EPLM), the sign problem is drastically weaken in the low temperature comparing with the original EPLM.  
The $Z_3$-EPLM also approaches smoothly to the original EPLM except for the nonanalytical point $\mu =M$ in zero temperature. 
At $\mu =M$, the expectation values of the Polyakov line( and its conjugate) has  different values in two models due to the existence or nonexistence of $Z_3$-symmetry.  
The Polyakov line can detect the symmetry structure of the zeros of the microcanonical probability density function, while the other quantities are insensitive to the structure. 
This property is not changed by varying $M/T$, since $\mu =M$ is the fixed point in the flavor symmetric EPLM. 
However, the effects of the flavor symmetry breaking and the spatial momentum of quarks may break this property and the expectation value of the Polyakov line may vanish even in the original EPLM when $M/T\to \infty$. 

The effects of the imaginary chemical potential $i\theta T$ at low temperature was also studied. 
The physical quantities at finite $\theta$ coincides with those at $\theta =0$ except for the neighborhood of ${\rm Re}(\mu )=M$. 
Hence the imaginary parts of the physical quantities can be induced only in the neighborhood. 
In the neighborhood of ${\rm Re}(\mu )=M$, the real parts of the number density and the modified Polyakov line vibrate violently. 
Large quark number density can be induced at the singular point. 
The $\theta$-dependence of the imaginary part of the physical quantities at the nonanalytical point is affected by the symmetry structure of the microscopic probability density function.  

It seems that the $Z_3$-symmetrized theory is equivalent to the original one with larger mass at least except just on the nonanalytical point, when $M/T$ is large enough. 
The trivial sign problem is expected to be weak in the $Z_3$-symmetrized theory. 
Hence, to explore the low temperature property, we may use the $Z_3$-symmetrized theory with smaller $M/T$ instead of the original one.  
However, in LQCD, it is known that there is a nontrivial hard problem on early onset of quark number density at zero temperature. (See, e.g., Refs.~\cite{Cohen,Nagata:2012ad} and references therein. ) 
Since $Z_3$-QCD is expected to approach to the original QCD in zero temperature limit, this problem may also happen in the $Z_3$-QCD when $M/T$ is very large. 
However, the problem may not occur just below $T_c$ in Fig.~\ref{phase-diagram} and we may use the probability density function in that region as the approximate probability density function to analyze the low temperature physics in the original QCD. 
Hence, the research of the lattice $Z_3$-QCD at the intermediate temperature may be important. 
Such research is now in progress.

\noindent
\begin{acknowledgments}
The authors are thankful to A. Miyahara, M. Ishii, J. Takahashi and M. Yahiro for fruitful discussions. 
This work is supported in part by Grant-in-Aid for Scientific Research (No. 18K03618 and 20K03974) from Japan Society for the Promotion of Science (JSPS). 
\end{acknowledgments}

%\clearpage

\appendix

%%%%%%%%%%%%%%%%%%%%%%%%%%%%%%%%%%%%%%%%%%%%%%%%
\section{Phase quenched approximation and phase factor}
\label{Aadd1}
%%%%%%%%%%%%%%%%%%%%%%%%%%%%%%%%%%%%%%%%%%%%%%%%

One of the simple approaches to the sign problem is the reweighting method.   
In this method, one can calculate the expectation value $\langle O \rangle^\prime$ of the quantity $O$ with 
a approximate weighting function $F^\prime (U)$ which is real and nonnegative.  
%%%%%%%%%%%%%%%%
\begin{eqnarray}
\langle O \rangle^\prime =\int {\cal D} U O(U) {F^\prime (U)\over{Z^\prime}};~~~~~
Z^\prime &=&\int {\cal D} U F^\prime (U). ~~~~~ 
\label{Z_QCD_d}
\end{eqnarray}
%%%%%%%%%%%%%%%%
where $U$ is the dynamical variables such as $\varphi_{c,\vix}$ in EPLM. 
 The true expectation value $\langle O \rangle$ is given by 
%%%%%%%%%%%%%%%%
\begin{eqnarray}
\langle O \rangle
=\int {\cal D} U O(U){F(U)\over{Z}}={\langle O {F(U)\over{F^\prime (U)}}\rangle^\prime \over{W}}, 
\label{RW}
\end{eqnarray}
%%%%%%%%%%%%%%%%
where $Z$ and $W=Z/Z^\prime $ are the true grand canonical partition function and the reweighting factor.       
When $W$ is very small, the true expectation value has large errors due to the division by $W$ in (\ref{RW}).  
In actual calculations, the phase-quenched function  
%%%%%%%%%%%%%%%%
\begin{eqnarray}
F^\prime (U)=|F^\prime (U)|, 
\label{PQ_QCD}
\end{eqnarray}
%%%%%%%%%%%%%%%%
is often used. 
This reweighting method is PQRW. 
In PQRW, $W$ is also called as "phase factor".

%%%%%%%%%%%%%%%%%%%%%%%%%%%%%%%%%%%%%%%%%%%%%%%%%%%%%%%%%%%%%%%%%%%%
\section{Analytical representation of physical quantities in EPLM at $\kappa =0$ with periodic boundary condition}
\label{A1}
%%%%%%%%%%%%%%%%%%%%%%%%%%%%%%%%%%%%%%%%%%%%%%%%%%%%%%%%%%%%%%%%%%%%

In the three flavor EPLM at $\kappa =0$ with periodic boundary condition, the grand canonical partition function is given by
%%%%%%%%%%%%%%%%%
\begin{eqnarray}
Z_{\rm PB}=\prod_{l=0}^D\left[ \int_{-\pi}^\pi du \int_{-\pi}^\pi dv e^{-2^l{\cal L}(u,v)}\right]^{~_D\text{C}_l(L_{\rm s}-2)^{D-l}}, 
\label{Z_PB}
\end{eqnarray}
%%%%%%%%%%%%%%%%
where $D$ is the dimension of the spatial space and  $L_{\rm s}^3(=N_{\rm s})$ is the number of the lattice spatial sites.    
Note that 
%%%%%%%%%%%%%%%%
\begin{eqnarray}
\sum_{l=0}^D~_D\text{C}_l(L_{\rm s}-2)^{D-l}~2^l=(L_{\rm s}-2+2)^D=L_{\rm s}^D=N_{\rm s}, 
\nonumber
\label{binomial}
\end{eqnarray}
%%%%%%%%%%%%%%%%
is satisfied. 

Similarly, the partition function for an approximate Lagrangian ${\cal L}^\prime$, 
the pressure, the quark number density, the scalar density, the averaged values of the Polyakov line and its conjugate at $\kappa =0$ are given by   
%%%%%%%%%%%%%%%%%
\begin{eqnarray}
Z_{\rm PB}^\prime =\prod_{l=0}^D\left[ \int_{-\pi}^\pi du \int_{-\pi}^\pi dv e^{-2^l{\cal L}^\prime (u,v)}\right]^{~_D\text{C}_l(L_{\rm s}-2)^{D-l}}, 
\label{Z_PB_d}
\end{eqnarray}
%%%%%%%%%%%%%%%%
%%%%%%%%%%%%%%%%%
\begin{eqnarray}
{\cal P}_{\rm PB}&=&T\sum_{l=0}^D{~_D\text{C}_l(N_{\rm s}-2)^{D-l}\over{L_{\rm s}^D}}
\nonumber\\
&\times&\log{\left[ \int_{-\pi}^\pi du \int_{-\pi}^\pi dv e^{-2^l{\cal L}(u,v)}\right]}, 
\nonumber\\
\label{P_PB}
\end{eqnarray}
%%%%%%%%%%%%%%%%
%%%%%%%%%%%%%%
\begin{eqnarray}
n_{\rm q~PB}&=&\sum_{l=0}^D{2^l~_D\text{C}_l(L_{\rm s}-2)^{D-l}\over{L_{\rm s}^D}}
\nonumber\\
&\times&{\int_{-\pi}^\pi du \int_{-\pi}^\pi dv \left(-T{\partial L\over{\partial \mu}} \right) e^{-2^l{\cal L}(u,v)}\over{\int_{-\pi}^\pi du \int_{-\pi}^\pi dv e^{-2^l{\cal L}(u,v)}}}, 
\label{n_PB}
\end{eqnarray}
%%%%%%%%%%%%%%%%
%%%%%%%%%%%%%%
\begin{eqnarray}
n_{\rm s~PB}&=&\sum_{f=u,d,s}\sum_{l=0}^D{2^l~_D\text{C}_l(L_{\rm s}-2)^{D-l}\over{L_{\rm s}^D}}
\nonumber\\
&\times&{\int_{-\pi}^\pi du \int_{-\pi}^\pi dv \left(T{\partial L\over{\partial M_f}} \right) e^{-2^l{\cal L}(u,v)}\over{\int_{-\pi}^\pi du \int_{-\pi}^\pi dv e^{-2^l{\cal L}(u,v)}}}, 
\label{ns_PB}
\end{eqnarray}
%%%%%%%%%%%%%%%%
%%%%%%%%%%%%%%
\begin{eqnarray}
\langle P_\vix \rangle_{\rm PB}&=&\sum_{l=0}^D{2^l~_D\text{C}_l(L_{\rm s}-2)^{D-l}\over{L_{\rm s}^D}}
\nonumber\\
&\times&{\int_{-\pi}^\pi du \int_{-\pi}^\pi dv P_\vix (u,v)  e^{-2^l{\cal L}(u,v)}\over{\int_{-\pi}^\pi du \int_{-\pi}^\pi dv e^{-2^l{\cal L}(u,v)}}}, 
\label{Phi_PB}
\end{eqnarray}
%%%%%%%%%%%%%%%%
%%%%%%%%%%%%%%
\begin{eqnarray}
\langle P_\vix^* \rangle_{\rm PB}&=&\sum_{l=0}^D{2^l~_D\text{C}_l(L_{\rm s}-2)^{D-l}\over{L_{\rm s}^D}}
\nonumber\\
&\times&{\int_{-\pi}^\pi du \int_{-\pi}^\pi dv P_\vix^*(u,v) e^{-2^l{\cal L}(u,v)}\over{\int_{-\pi}^\pi du \int_{-\pi}^\pi dv e^{-2^l{\cal L}(u,v)}}}. 
\label{Phi_PB_conj}
\end{eqnarray}
%%%%%%%%%%%%%%%%
In this case, not only the phase factors, but also the other physical quantities depend on $N_{\rm s}$. 
However, it can be easily seen that the effects of the boundary conditions vanish and   
the $N_{\rm s}$-dependences of these thermodynamical quantities also vanish in the limit of $N_{\rm s}\to \infty$. 

%\newpage

%%%%%%%%%%%%%%%%%%%%%%%%%%%%%%%%%%%%%%%%%%%%%%%%%%%%%%%%%%%%%%%%%%%%%%%%%%%%%%%%%%%%% References
%%%%%%%%%%%%%%%%%%%%%%%%%%%%%%%%%%%%%%%%%%%%%%%%%%%%%%%%%%%%%%%%%%%%%%%%%%%%%%%%

\end{document}